\documentclass{ametsocV6.1}

\usepackage{color}
\usepackage{amsmath} 
\usepackage{amssymb}
\usepackage{ragged2e} 
\usepackage{syntonly}  
\usepackage{float}
\usepackage{graphicx}
\usepackage{caption}
\usepackage{subfig}
\usepackage{footnote}
\usepackage{varwidth}
\usepackage{multirow}
\usepackage{fancyhdr}
\usepackage{extramarks}
\usepackage{soul}
\usepackage{array}
\usepackage{natbib}
\usepackage{url}
\usepackage{hyperref}

\bibpunct{(}{)}{;}{a}{}{,}

\hypersetup{colorlinks,%
	citecolor = blue,%
	filecolor  = blue,%
	linkcolor = blue,%
	urlcolor  = blue,%
	pdftex}

\pdfoutput=1

\makesavenoteenv{tabular}
\makesavenoteenv{table}

\newcolumntype{L}[1]{>{\raggedright\let\newline\\\arraybackslash\hspace{0pt}}m{#1}}
\newcolumntype{C}[1]{>{\centering\let\newline\\\arraybackslash\hspace{0pt}}m{#1}}
\newcolumntype{R}[1]{>{\raggedleft\let\newline\\\arraybackslash\hspace{0pt}}m{#1}}

\title{Prediction of Drought and Flash Drought in Africa at the Seasonal-to-Subseasonal Scale using the Community Research Earth Digital Intelligence Twin Framework}

\authors{Stuart Edris\correspondingauthor{Stuart Edris, sgedris@ou.edu}\aff{a},
  Amy McGovern\aff{a, b}, 
  and Jason Hickey\aff{c}} 

\affiliation{\aff{a}{School of Meteorology, University of Oklahoma}\\
\aff{b}{School of Computer Science, University of Oklahoma}
\aff{c}{Google Research Africa}
}

\nolinenumbers
\fancypagestyle{titlepage}
{
    \pagestyle{myheadings}
    \pagestyle{plain}
    \pagestyle{fancy}
    \fancyhf{}
    \cfoot{\thepage}
    \chead{Submitted to: Artificial Intelligence for Environmental Sciences (AIES)}
}

\abstract{Droughts and flash droughts (rapidly developing droughts; FDs) remain impactful events that are known to desiccate landscape and destroy crops. In particular, droughts in Africa are often more impactful than in other locations, such as the United States or Europe, due to many regions in Africa heavily depending on local agriculture for sustenance. In recent years, large machine learning (ML) models, such as GraphCast and AIFS, have emerged as effective tools for global weather prediction. However, sparse data observations and few ML studies in Africa have left it unclear if these ML models retain their skill when focused on Africa. As such, this project seeks to examine the predictability of drought and FD in Africa using a CrossFormer model based on the Community Research Earth Digital Intelligence Twin (CREDIT) framework developed by NSF NCAR. Our CrossFormer model, termed DroughtFormer, incorporates variables from the ERA5 and GLDAS2 reanalyses and the IMERG and MODIS satellite observations, and employs dry air mass and moisture conservation, to predict soil moisture, vegetation health, and other drought-related surface variables. While DroughtFormer displayed lower accuracy in predicting precipitation and FD indices, it showed significant skill in predicting the remaining variables, delivering stable and skillful forecasts out to 90-day lead times (either beating out or having comparable skill to climatology). In particular, DroughtFormer skillfully represented climate anomalies for key variables, such as soil moisture (though it struggled with the magnitude of the anomalies). Thus, DroughtFormer showed significant promise in representing and predicting agricultural level drought in a region that is heavily impacted by drought events.}

\begin{document}
    \maketitle

    \statement
     Drought remains an impactful climate extreme, particular in Africa where droughts are more severe due to a lack of irrigation and drought mitigation strategies. This leads to a reliance on monsoonal cycles for water within Africa, and is thus sensitive to weak monsoons or monsoonal failures. In addition, with advances in machine learning (ML) models, new deep learning models have been developed that show considerable skill in predicting weather for several days out. However, these studies have been global in scale, and have not given any attention to drought, nor focused on Africa, nor other data sparse regions, despite the need for improved weather predictions over the continent. To bridge this gap, we developed a new machine learning model, named DroughtFormer, for the purpose of drought prediction with emphasis of drought prediction over Africa. DroughtFormer is capable of producing stable, skillful predictions of drought related variables at the daily timescale out to 90 days, encompassing the full seasonal-to-subseasonal (S2S) timescale. In addition, DroughtFormer produced useful forecasts of climate anomalies for those same variables, correctly forecasting the timing, location, and sign of the climate anomalies (thus indicating the presence of drought multiple weeks in advance). These capabilities of DroughtFormer make it a useful decision-support tool for drought prediction, especially with its ability to deliver useful information at lead times currently unavailable in existing numerical weather prediction (NWP) methods. The main limitations of DroughtFormer was found to be that it struggled to correctly predict precipitation and the magnitudes of the climate anomalies. Failure to capture climate anomaly magnitudes also led to limitations in forecasting rapid intensification of drought conditions. Overall, DroughtFormer proved a useful decision-support tool, in being able to deliver reliable predictions across the entire S2S range, a time range that is currently not provided with from existing NWP forecasts or are heavily reliant on climatology, and in detecting the presence of drought, delivering a new baseline for deterministic forecasts in Africa for numerous variables.

    \section{Introduction}
    Droughts characterize dry extremes in existing moisture conditions, often understood as a shortage in available water for flora, fauna, and people within an ecosystem. This shortage causes disruptions in the ecosystem, such as loss of vegetation, loss of crop and livestock, and droughts can, in worst case scenarios, result in famine (e.g., crop damages in the central United States in 2012;~\citealt{Basara_2019}, health impacts in Russia in 2010;~\citealt{Christian_2020}, and drought induced famines in eastern Africa 2016 -- 17 and 2021 -- 2023;~\citealt{Funk_2018, Palmer_2023, Kimutai_2025, undrr_2024}). Dry conditions in different moisture variables describe different levels of drought. For example, below normal precipitation describing short-term meteorological drought, soil moisture deficits describes monthly to seasonal agricultural drought, and groundwater deficit describes seasonal to annual and multiannual hydrologic drought~\citep{Wang_2018}. In general, drought is often seen as a slow event that can last for multiple years or longer, yielding prolonged impacts.
    \par

    However, equally prominent is the occurrence of rapidly evolving drought conditions, termed flash drought (FD), that results in large impacts on vegetation and crops due to the rapid loss of available moisture. FDs particularly develop on the order of 2 - 6 weeks depending on the method used for defining FD~\citep{Otkin_2018, Lisonbee_2021}. This has the effect that FD occurs on the seasonal-to-subseasonal (S2S) timescale, presenting a difficult forecasting challenge, amplified by the fact that FD development is characterized by complex surface interactions~\citep{Lisonbee_2021, Tyagi_2022}. However, previous works have found that the key drivers for FD are evaportranspiration (ET), potential ET (PET), and soil moisture (SM)~\citep{Tyagi_2022, Osman_2022, Mukherjee_2022a, Rakkasagi_2023}, though other variables such as precipitation are often used as the deficit in precipitation can often initiate the process of FD development~\citep{Noguera_2021, Tyagi_2022, Rakkasagi_2023}. In addition, ratios of ET and PET have proven useful in numerous studies to evaluate vegetation health and increasing evaporative stress, while evolving on the same timescale as FD~\citep{Dale_2017, Christian_2019b, Gou_2022}. In order to compensate for sparse measurements of ET and PET in space and time, ratios of ET to PET are normally obtained via satellite (\citealt{Ford_2023}), or via reanalysis models, to obtain the standardized evaporative stress ratio (SESR;~\citealt{Christian_2019b, Gou_2022}). Like ET and PET SM is often estimated from satellite observations or reanalysis models to compensate for sparse measurements (e.g.,~\citealt{Christian_2022, Osman_2022, Otkin_2021, Sehgal_2021, Mukherjee_2022a, Mukherjee_2022b, Yuan_2019}).
    \par
   
    In addition, the continent of Africa is worth particular note in terms of impacts, as droughts are often more impactful on the African continent than other locations. Eastern Africa, for example, has seen multiple drought induced famines in the last 15 years~\citep{Funk_2018, Adloff_2022, Doi_2022, Palmer_2023, Kimutai_2025}. Yet despite this, Africa is often under represented in drought studies~\citep{Vicente_Serrano_2012, Dikshit_2022a}. In addition, while there have been global FD studies that have discussed FD in Africa (e.g., the Sahel and Great Rift Valley were found to be regions that experience more frequent FDs;~\citealt{Christian_2021, Yuan_2023}), there have been only one study to our knowledge that have focused on FD in Africa~\citep{Anande_2026}. Since there is little irrigation within Africa, most of the droughts on the continent have been connected to precipitation deficits~\citep{Verdin_2005, Naumann_2012, Guido_2020, Ayugi_2022}, though some studies have incorporated SM to help gauge vegetation health during drought~\citep{Shukla_2014, Kew_2021, Wu_2021}. This, in turn, makes understanding the precipitation climatology and teleconnections more important for understanding and predicting drought and FD in Africa. In particular, since Africa straddles the Equator, most of sub-Saharan Africa does not experience traditional extratropical seasons, but annual monsoonal, or monsoon-like, rainy seasons associated with the migration of the inter-tropical convergence zone (ITCZ), and dry seasons~\citep{Liebmann_2012, Maidment_2014, Tarnavsky_2014, Wainwright_2021}. Notably, some portions of Africa, such as southern Somalia, Kenya, Tanzania, and the Congo Basin experience two rainy seasons as the ITCZ moves northward in March to July and as the ITCZ migrates south in September to November~\citep{Hoell_2013, Liebmann_2014, Lyon_2014, Nicholson_2017, Jiang_2021}. Droughts then occur when the rainy season underperforms or fails. For locations with two rainy seasons, the first rainy season in March to June is typically the more reliable and delivers higher rainfall amounts, and underperformance then can yield significant droughts~\citep{Adloff_2022}. Variations in African precipitation is often connected with SSTs in the Indian and Pacific Oceans (upstream in the normal Walker Circulation), namely the El Nino Southern Oscillation (ENSO) and Indian Ocean Dipole (IOD)~\citep{Nicholson_1997, Misra_2003, Saji_2003, Vicente_Serrano_2012, Tierney_2013, Hoell_2013, Liebmann_2014, Lyon_2014, Nicholson_2015, Nicholson_2017, Wainwright_2020, Jiang_2021}.
    \par
   
    In recent years, machine learning (ML) has become increasingly prominent in its applications towards weather prediction (e.g.,~\citealt{Nafii_2022, Lam_2024, Nipen_2024, Schreck_2025_CREDIT}). Among other weather prediction applications, ML has also been used to predict drought, often in the form of predicting drought indices such as the Standardized Precipitation Index (SPI;~\citealt{AghaKouchak_2022, Dikshit_2022b}). ML studies in representing and predicting drought in Africa have been conducted, with some applications towards groundwater~\citep{Thomas_2020b_DRIP, Fankhauser_2022, Holland_2023}, precipitation~\citep{Deman_2022} and other drought indices~\citep{Nafii_2022, Ferchichi_2024}. However, in comparison to progress in general numerical weather prediction (NWP), advances in drought prediction have lagged behind, often focusing on traditional ML (e.g., random forests and XGBoosting) methods or simple neural networks. For NWP in general, large and sophisticated deep learning models have been developed that have been able to make predictions of global weather variables that are comparable in skill to, if not better than, operational NWP methods~\citep{Lam_2024, Nipen_2024, Schreck_2025_CREDIT}. In particular, development of global models such as GraphCast~\citep{Lam_2024}, AIFS~\citep{Lang_2024}, Bris~\citep{Nipen_2024}, and FastCast~\citep{Dunstan_2025} have shown the skill of graph neural networks (GNNs) in NWP up to 5+ day lead times (predictions made in 6 hour intervals), identifying GNNs as one of the more skillful types of neural networks for NWP alongside different types vision transformers (e.g., Pangu and FuXi;~\citealt{Bi_2023, Chen_2023}). In addition, CrossFormer networks, a mixture of vision transformer and convolutional U-networks, has also been shown to have high forecast skill compared to pure vision transformers~\citep{Schreck_2025_CREDIT}. 
    \par
    
    Of particular note was the release of the Community Research Earth Digital Intelligence Twin (CREDIT) framework~\citep{Schreck_2025_CREDIT}, 
    which was designed to provide a framework on which to create, train, evaluate, and deploy ML models. The CREDIT framework provides and end-to-end pipeline, built in PyTorch, to facilitate the development of ML model for earth sciences, as well as provide access to multiple state-of-the-art, adjustable ML architectures that are capable of running on High-Performance Computing (HPC) with multiple GPUs without requiring expertise in parallel processing or requiring the tens to hundreds of GPUS needed to train networks like GraphCast or AIFS. Thus, the release of this framework enables us to investigate drought with a state-of-the-art ML model capable of delivering skillful global forecasts. This is particularly unique as large-scale examination of drought forecast, focusing more on driving variables rather than indices, are more limited in number, and even more so on Africa (to our knowledge there is only one ML and drought study focusing on Africa as a whole,~\citealt{Ferchichi_2024}, but most only focus on specific subregions, such as part of Kenya).
    \par
   
    Despite advances in ML prediction techniques, the use of ML in predicting FD is still new, with relatively few studies (e.g.,~\citealt{Zhang_2022, Foroumandi_2024}). Of these, all have focused on regions typical for FD studies, such as the United States and China. To our knowledge, there has only been one study~\citep{Anande_2026} that focuses on FD in Africa (existing knowledge on FD in Africa stems from global studies), and none on FD prediction in Africa. In addition, prediction of standard droughts within Africa still requires improvement, given the severity of impacts droughts can have on the continent. As such, this study seeks to bridge this gap by exploring the use of a large, complicated CrossFormer model, developed under the CREDIT framework to predict drought and FD on the S2S timescale (that is, the time scale on which FD and short-term drought occurs). The goals of this paper are: (1) To predict variables related to drought (temperature and 30 day precipitation accumulation), FD (ET, PET, and SM at multiple depths), FD indices (SESR and FD intensity index for multiple depths;~\citealt{Otkin_2021}), and indices related to vegetation health (Normalized Differential Vegetation Index, Enhanced Vegetation Index, Leaf Area Index, and fraction of absorbed Photosynthetically Active Radiation). We also evaluate if a state-of-the-art ML model is capable of performing on a similar scale to, or outperform, climatology predictions in Africa (given the difficulty in S2S predictions, and the standard drop off in skill after a few weeks;~\citealt{Olaniyan_2025, Endris_2021, Phakula_2024}). (2) Given there are so few studies focused on the African continent, the second goal is to set a benchmark performance for deterministic ML prediction of drought in Africa which future studies can compare against and improve upon. (3) Identify current limitations of ML prediction of drought, and ways in which to improve ML prediction as compared to other ML NWP studies, particularly for the S2S timescale. 
    \par
    
    The remaining sections will detail the data and CrossFormer model, named DroughtFormer, (Sec.~\ref{sec:data_methods}), CrossFormer predictions on the S2S timescale (Sec.~\ref{sec:S2S_pred}), and a prediction case study (Sec.~\ref{sec:case_study}), before summarizing the results and concluding (Sec.~\ref{sec:conclusion}).

    \section{Data and Methods}\label{sec:data_methods}
    \subsection{Data}
    Our model synthesized data from multiple sources, with data from both satellite and reanalysis datasets. Given the sparsity of observations in Africa, remote-sensing satellites are often taken as the closest form of comprehensive observations available for the continent (e.g.,~\citealt{Liebmann_2012, Naumann_2012, Lyon_2014, Sheffield_2014, Maidment_2017, Macharia_2022}). As such we incorporated observations taken from the MODIS dataset, to characterize vegetation health~\citep{Myneni_2021_MODIS, Running_2021_MODIS, Fruedl_2022_MODIS, Lyapustin_2023_MODIS} and IMERG, which has been shown to accurately capture precipitation amounts~\citep{Olson_2019-IMERG_land-sea_data, Huffman_2023-IMERG_data}. MODIS reflectance data was collected from both terra and aqua satellites, which was used to calculate the Normalized Differential Vegetation Index (NDVI) and Enhanced Vegetation Index (EVI). MODIS data was also collected for  Leaf Area Index (LAI) and fraction of absorbed Photosynthetically Active Radiation (fPAR). MODIS delivers its data in 8-day periods, corresponding to the orbit of the satellites. For our study, MODIS  data was interpolated to the daily time scale using the most recent observation to gap fill. IMERG merges precipitation observations from a constellation of satellites observations to create a global observation network of precipitation, with multiple latencies (and thus time scales). In particular, IMERG offers global precipitation datasets interpolated with observations down to the hourly time scale, which can be used to obtain daily accumulation of precipitation (the final product we used). 30 day accumulated precipitation was also obtained to forecast the seasonal rainfall that occurs within Africa. 30 day accumulations were obtained by applying a 30 day, centered running mean to the daily precipitation data.
    \par

    Surface variables were collected from the Global Land Data Assimilation System, version 2 (GLDAS2~\citealt{Rodell_2004_GLDAS,Beaudoing_2020_GLDAS}) and the European Centre for Medium-Range Weather Forecasts' (ECMWF) global ERA5 reanalysis dataset~\citep{Hersbach_2023}. The ERA5 reanalysis dataset provides a detailed $0.25^{\circ}\times 0.25^{\circ}$, hourly data for multiple global variables and numerous isobaric heights from 1950 until the present. GLDAS2 provides detailed 3 hourly analysis (from 2000 to present) of global variables at the $0.25^{\circ}\times 0.25^{\circ}$ resolution, with focus on accurate representation of surface variables. Any additional variables not provided by GLDAS2, such as dewpoint and upper-air variables, were taken from ERA5. In addition, any missing data in the GLDAS2 (e.g., below 60$^{\circ}$S) was replaced with the ERA5 variables. Upper-air variables incorporated into the model were those shown to impact precipitation variability via teleconnection and synoptic patterns (i.e., u- and v-wind components and geopotential at 500 mb and 200 mb). Specific humidity and cloud liquid ice and water content at 500 mb and 200 mb was also incorporated to enforce moisture conservation (see Section 2\ref{sec:crossformer_model}). Lastly, the climate indices most important in driving long-term drought and FD (ENSO and IOD;~\citealt{Saji_2003}) were also collected from the COBE, ERSSTv5 and HadISST datasets~\citep{Rayner_2003_hadisst, ishii_2005_cobe, Huang_2017_ersst}. Other climate indices were explored but ultimately did not help DroughtFormer make accurate predictions (Appendix~\ref{sec:appendix_a}). The final list of variables, which includes 25 input variables and 23 outputs, along with which dataset the variable was derived from, can be found in Table~\ref{tab:list_of_variables}). 

    \begin{table}[]
        \caption{List of variables incorporated into the CrossFormer network. $^{\ddagger}$ indicates upper air variables (upper air variables at 200 mb and 500 mb were incorporated for this study). Variables marked with $^*$ were obtained from GLDAS2. Variables marked with $^{**}$ were obtained from HADISSTs and ERAsstv5. Variables marked with $^\dagger$ were obtained from IMERG. Variables marked with $^{\dagger\dagger}$ were obtained from MODIS. Variables without any of above markings were obtained from ERA5.}
   
        \hspace{-0.4cm}
        \small
        \begin{tabular}{|p{3.2cm}|p{3.0cm}|p{3.0cm}|p{3.0cm}|p{2.8cm}|}
            \hline
            \multicolumn{1}{|p{2.8cm}|}{Prognostic Variables (Input \& Output)} & \multicolumn{3}{c|}{Input Only} & \multicolumn{1}{p{2.8cm}|}{Diagnostic Variables (Output Only)} \\
            \hline
            & \multicolumn{1}{c|}{Dynamic Variables} & \multicolumn{1}{c|}{Forcing Variables\footnote{Forcing variables are cyclical/climatological; variables averaged in time to deliver the average value each date in the year}} & \multicolumn{1}{c|}{Static Variables\footnote{Static variables are unchanging in time, they are simply latitude by longitude format}} & \\
            \hline
            \small - U--Wind Component$^{\ddagger}$ & \small - ENSO$^{**}$ & \small - Net Surface Shortwave & \small - Land-Sea Mask & \small - 30 Day Precipitation \\
         
            \small - V-Wind Component$^{\ddagger}$ & - IOD$^{**}$ & \small Radiation Climatology$^*$ & \small - High Vegetation Type & \small Accumulation$^\dagger$ \\

            \small - Geopotential$^{\ddagger}$ & \small - Net Surface Shortwave & & \small - High Vegetation Cover & \small - NDVI$^{\dagger\dagger}$ \\

            \small - Total Specific Humidity & \small Radiation$^*$ & & \small - Low Vegetation Type & \small - EVI$^{\dagger\dagger}$ \\

            \small ($q_{tot}$)$^{\ddagger}$\footnote{$q_{tot} = $ specific humidity + total cloud liquid water content + total cloud ice water content} & \small - Surface Wind Speed$^*$ & & \small - Low Vegetation Cover & \small - LAI$^{\dagger\dagger}$ \\

            \small - Temperature$^*$ & \small - Surface Wind Gusts & & & \small - fPAR$^{\dagger\dagger}$ \\

            \small - Dewpoint Temperature & & & & \small - SM Layer 1$^*$\footnote{\label{SM_depths}Soil Moisture (and FDII by extension) layer depths for GLDAS2 are as follows. Layer 1: 0 – 10 cm, layer 2: 10 – 40 cm, layer 3: 40 – 100 cm, layer 4: 100 – 200 cm} \\

            \small \raggedright - 1 Day Precipitation & & & & \small - SM Layer 2$^*$\footref{SM_depths} \\

            \small \raggedright Accumulation$^\dagger$ & & & & \small - SM Layer 3$^*$\footref{SM_depths} \\
         
            \small - Surface Pressure$^*$ & & & & \small - SM Layer 4$^*$\footref{SM_depths} \\
         
            \small - Evaporation$^*$ & & & & \\

            \small - Potential Evaporation$^*$ & &  & & \\
            \hline
        \end{tabular}
    
        \label{tab:list_of_variables}
    \end{table}
    \par

    Given its wide use and ability to represent vegetation health under evaporative stress, we explored the use of the Standardized Evaporative Stress Ratio (SESR), calculated using ET and PET according to~\cite{Christian_2019b}. In addition, a quantitative approach to FD was also incorporated using the Flash Drought Intensity Index (FDII), which uses SM to quantify the intensity of rapid onset of drought, and the severity of the drought~\citep{Otkin_2021}. Given multiple SM depths were incorporated for this study, the FDII was then calculated for each depth. Originally, the indices were incorporated into DroughtFormer as a diagnostic variable, however this proved problematic for the model and they were ultimately calculated from the model predictions (See Sec.~\ref{sec:S2S_pred}).
   
    For each variable, data was collected from 2001 (the earliest complete year of satellite observations) to 2024, providing 24 years of data in total. We partitioned the 24 years of multi-source satellite and reanalysis data into training (2001 -- 2017; 2022 -- 2024), validation (2018 -- 2019), and testing (2020 -- 2021) sets. The data was also limited to the most coarse dataset, in this case ERA5 and GLDAS2 at 0.25$^{\circ}$ $\times$  0.25$^{\circ}$ spatial resolution. However, it was found that using quarter degree resolution verses 1$^{\circ}$ $\times$ 1$^{\circ}$ resolution did not noticeably improve performance or predictions (in fact DroughtFormer had a more difficult time learning long lead times) and quarter degree resolution required more than an order of magnitude of computation and training time, thus we averaged the datasets down to the $1^{\circ} \times 1^{\circ}$ resolution.

    \subsection{DroughtFormer}\label{sec:crossformer_model}
\begin{figure}
        \noindent\fbox{\includegraphics[width=1.0\textwidth,angle=0]{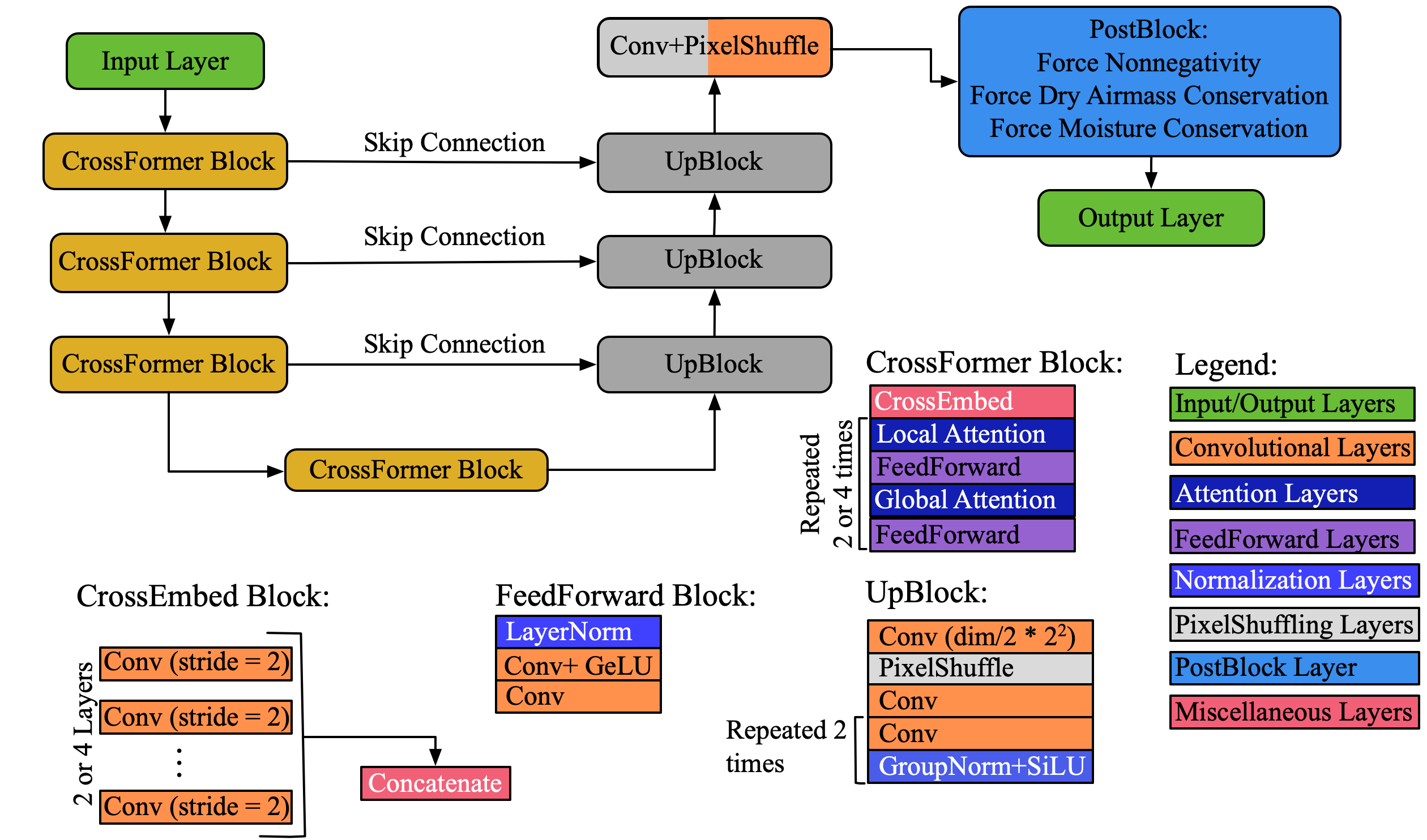}}\\
        \caption{Illustration of the DroughtFormer model. }\label{fig:f1_crossformer_illustration}
    \end{figure}

    DroughtFormer, a customized CrossFormer model, was built on the CREDIT framework~\citep{Schreck_2025_CREDIT}, which in turn built its neural network models in PyTorch. CrossFormer models combine the typical structure of a convolutional U-network with multiheaded attention used in transformer networks to learn complex spatiotemporal patterns (Fig.~\ref{fig:f1_crossformer_illustration}). Thus, the model can be split into an encoder, which processes and downsizes the data into latent space, and a decoder that up samples the data onto the original map. The core of the CrossFormer is found in the encoder, which uses four CrossFormer blocks. Each CrossFormer block in the model starts with a cross-embedding layer, which is a set of 4 convolutional layers (in parallel) with multiple kernel sizes of $4\times4$, $8\times8$, $16\times16$, and $32\times32$ for each convolutional layer~\citep{Schreck_2025_CREDIT}. Downsampling was also performed in this embedding layer. Long and short distance multiheaded attention layers are then used to encode and process the data in each CrossFormer block. Short distance attention operates on $N\times N$ neighborhoods ($5\times5$ for our model), while long distance attention operates on fixed intervals to facilitate long-range interactions (\citealt{Schreck_2025_CREDIT}; for our model, the global attention size was 2, 2, 2, and 1 for each respective CrossFormer block). The decoder for our model differs from the WxFormer in~\cite{Schreck_2025_CREDIT}, in that we used pixel shuffling to up sample instead of convolutional transpose. Pixel shuffling, a method that up samples via a style of data re-arrangement~\citep{Shi_2016}, was found to improve model stability and reduce artifacts found at longer lead times (see Appendix~\ref{sec:appendix_a}). Skip connections were applied to connect each CrossFormer block (except the last, which acts as a processor step between the encoder and decoder) with corresponding up sample blocks to improve model learning and stability. Spectral normalization and mirror boundary padding~\citep{Schreck_2025_CREDIT} were also used to improve model stability and performance.

    Additionally, the CREDIT framework also incorporates the use of physics constraints~\citep{Watt-Meyer_2024, Chapman_2025, Sha_2025}. These constraints apply corrections to non-physical values (e.g., forcing negative values to zero for variables like precipitation), and adjust surface pressure, precipitation, and temperature predictions prior to loss calculation to enforce dry air mass, moisture, and energy conservation. The conservation schemes use ratios of mass, moisture, and energy variables to determine deviations from conserved dry air mass/moisture/energy in an atmospheric column (requiring certain variables, such as total specific humidity, to be included in DroughtFormer) and apply corrections to the corresponding variable (see~\citealt{Chapman_2025, Sha_2025} for details). For this study, dry air mass and moisture conservation was enforced to better represent the moisture variables for drought, while energy conservation was omitted because the temperature corrections were not necessary for the drought predictions. In addition to improving precipitation predictions, the addition of physical constraints were found to reduce model tendencies to deviate towards unrealistic predictions at longer lead times (Appendix~\ref{sec:appendix_a}). The physics constraints combined with the use of pixel shuffling layers allowed DroughtFormer model to make stable predictions for up to 90 day lead times, covering the breadth of the S2S time range.
    \par

    Overall, DroughtFormer had a similar architecture to the WxFormer in~\cite{Schreck_2025_CREDIT}, save for the use of pixel shuffling layers over convolutional transpose and applied physics constraints. Modifications to the model architecture were explored, but the setup of 4 to 32 attention heads were found to strike a balance between an accurate CrossFormer model without significantly increasing computation requirements (this corresponds to an attention dimensionality of 128 to 1,024). Regularization was also employed to reduce overfitting~\citep{Madakumbura_2021, Marcolongo_2022}, specifically L2 regularization with a L2 parameter of $10^{-5}$, as well as a dropout of 0.05 for the attention and feed forward layers. DroughtFormer was trained on 4 A100 GPUs (making its computation requirements cheaper than other large ML models, such as GraphCast and AIFS), and it was trained for a total of 12.2 days for 220 epochs (Table~\ref{tab:training_parameters}). Additional details and justification of the model parameters and how they were determined can be found in the Appendix~\ref{sec:appendix_a}.
    \par

    Model training was split into two parts. First, the model was pre-trained using single-step training for 120 epochs, where the model was trained on one day predictions to learn the initial conditions for each day. 120 epochs was chosen to ensure model learned initial conditions, and to give it sufficient time to converge (e.g., epoch 107 was the best performing epoch). The second part consisted of multistep training, where the model was trained on 3 days of rollout predictions for 100 epochs. Multistep training has been shown to improve model stability and accuracy, particularly in longer lead times found in S2S predictions~\citep{Schreck_2025_CREDIT, Chapman_2025, Guo_2024} at the cost of increasing training time (because the amount of work needed to compute the loss is increased by a factor of the training step size; Table~\ref{tab:training_parameters}). In particular, the CrossFormer used for this experiment was trained on 3 days of predictions as experimentation found increasing the number of predictions significantly increased computation time without a significant improvement in prediction accuracy (Appendix~\ref{sec:appendix_a}).
    \par

    For model evaluation, 90 day forecasts were made for each day in the test data (2020 -- 2021) and the resulting skill scores were averaged over each prediction. Model performance was examined via the anomaly correlation coefficient (ACC) and root mean square error (RMSE), as well as through case study examinations to help ensure realistic predictions without significant artifacts. Variation in the model output was also determined by the ratio of predictable components (RPC) in order to determine DroughtFormer's ability accurately gauge variability in the data~\citep{Scaife_2018, Brocker_2023}. The RPC compares ensemble forecasts to true labels, though it can be modified to compare the variance of one set of model predictions to true labels~\citep{Brocker_2023}. In particular, RPC is calculated as the ratio of the correlation coefficient $\rho$ (between the model predictions and true labels) and the model correlation coefficient $\rho_f$:
    \begin{equation}
        \label{eq:rpc}
        RPC = \frac{\rho}{\rho_f}; \quad \rho_f = \sqrt{\frac{\sigma^2_m}{\sigma^2_m + \sigma^2_{\varepsilon}}}
    \end{equation}
    where $\sigma^2_m$ is the variance in the model predictions and $\sigma^2_{\varepsilon}$ is the variance in model error (that is, the variance in true labels minus model predictions). This ratio gives an ideal scenario of RPC $=1$, in which case the model perfectly capture the true label's variance. When the RPC $<1$, the correlation between the model and true labels is lower than the model correlation, giving the predictable fraction of variance in the model is lower than expected~\citep{Scaife_2018}. Meaning an RPC $<1$ implies the model is over confident or under dispersive. Conversely, an RPC $>1$ yields an anomalous signal-to-noise ratio in the model. For this study, the RPC was used to compare the anomalies in predicted variables to anomalies in the true labels.
    \par
    
    Finally, it is worth noting DroughtFormer was trained on global scale data, and outputted global predictions. However, the focus of this study was on Africa, and so performance metrics were calculated over Africa, unless otherwise stated in specific figures.

    \section{S2S Scale Predictions}\label{sec:S2S_pred}
\begin{figure}
        \noindent\fbox{\includegraphics[width=1.0\textwidth,angle=0]{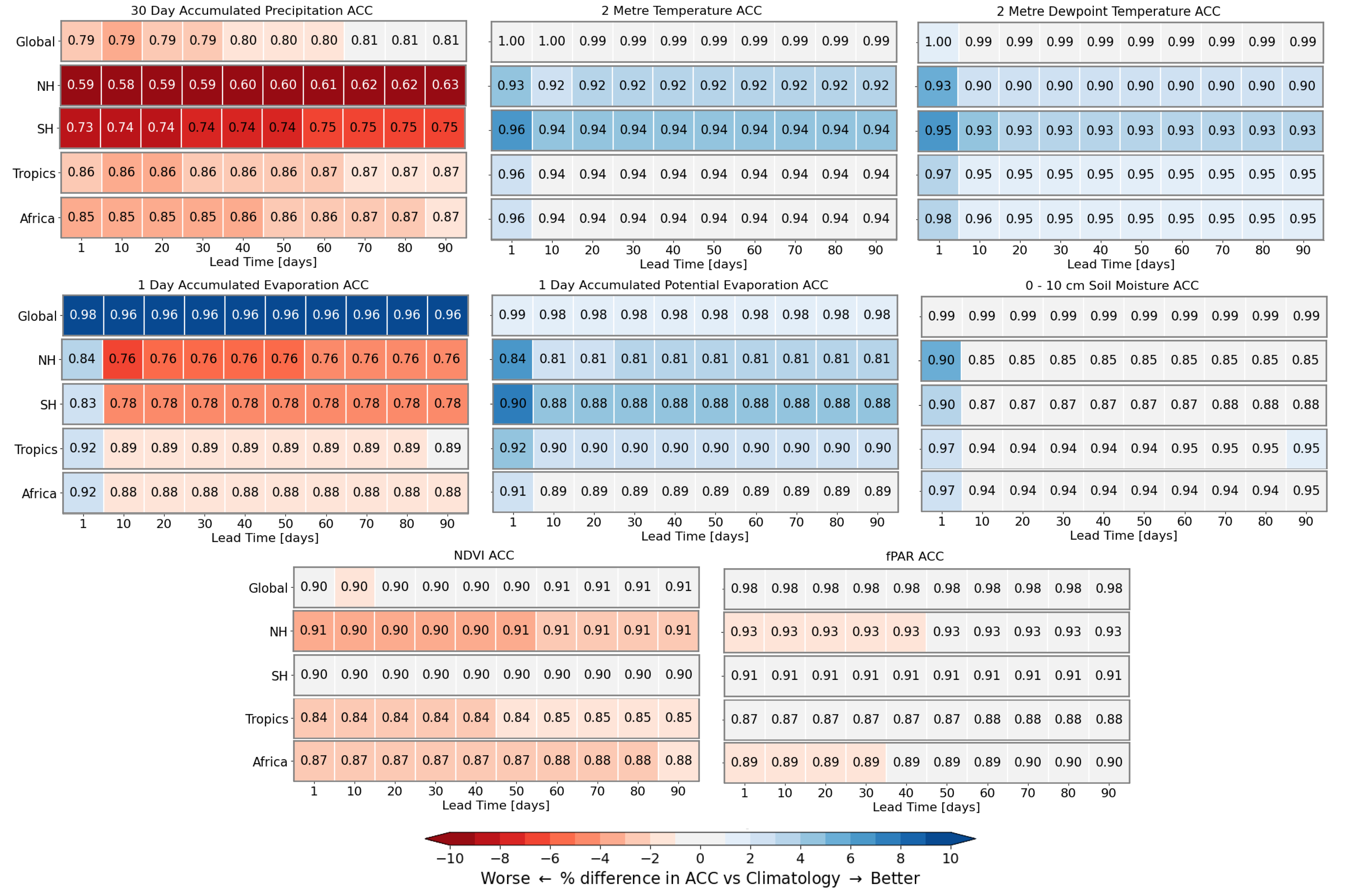}}\\
        \caption{Averaged ACC score cards over space and all forecasts in 2020 -- 2021 for 30 day accumulated precipitation, temperature, dewpoint temperature, ET, PET, near surface SM (0 - 10 cm), NDVI, and fPAR as a function of lead time. Colors indicate \% improvement/deterioration compared to the climatology forecast. Note global metrics includes sea grid points in the evaluation.}\label{fig:scorecard}
    \end{figure}

\begin{figure}
        \noindent\fbox{\includegraphics[width=1.0\textwidth,angle=0]{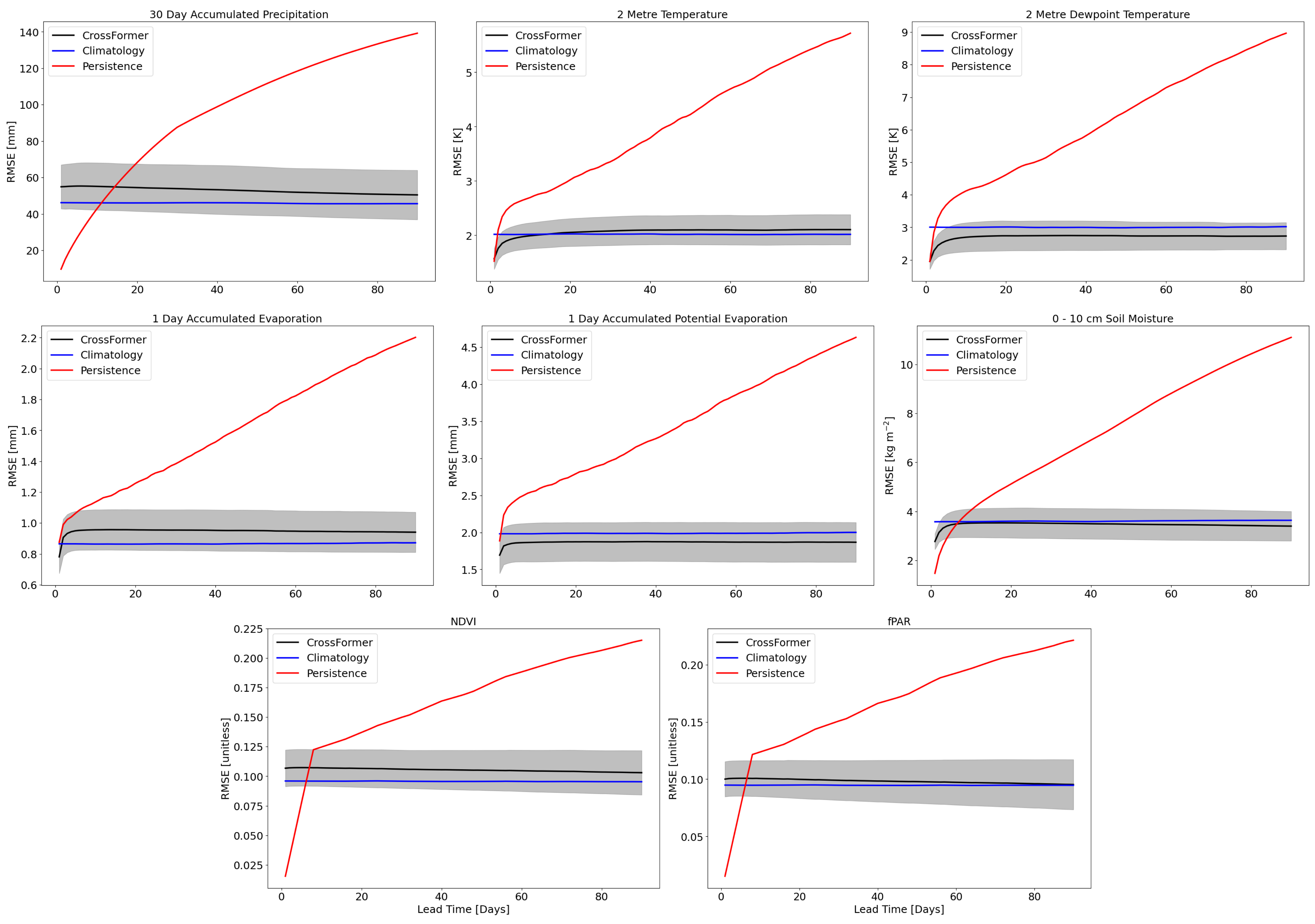}}\\
        \caption{Averaged RMSE over Africa for all forecasts from 2020 -- 2021 of 30 day accumulated precipitation, temperature, dewpoint temperature, ET, PET, near surface SM (0 - 10 cm), NDVI, and fPAR as a function of lead time, compared to persistence and climatology forecasts. Shading indicates 1 standard deviation in the CrossFormer predictive skill.}\label{fig:rmse_fig}
    \end{figure}

    For comparative analysis, DroughtFormer prediction skill was compared to climatology forecasts (that is, the 24 year average for a given grid point and day in the year) due to a lack of access to operational forecast methods used in Africa. Climatology was chosen as the baseline because, when compared to the overall means used in RMSE and ACC calculations, it produced higher skill scores than persistence, yielding a higher baseline for DroughtFormer to beat. DroughtFormer results showed a wide range of comparability with climatology in terms of traditional skill scores (Figs.~\ref{fig:scorecard} -~\ref{fig:rmse_fig}). Of particular note, DroughtFormer managed to maintain stable, skillful predictions out to 90 day lead times, encompassing the full S2S time scale. The ability of DroughtFormer to consistently deliver skillful predictions out to 90 days or further was found to be one of the main accomplishments of our model as many existing NWP methods struggle with this task. DroughtFormer's ability to deliver stable predictions for the full S2S time was attributed to the up sampling method (pixel shuffling), the physics constraints applied to the model (Appendix~\ref{sec:appendix_a}), and to the daily timescale examined in this study. That is, learning and predicting daily averages and accumulations was found to be an easier task than trying to learn diurnal cycles at the 6-hour timescale prevalent in most ML weather prediction studies. Additionally, pixel shuffling required a convolutional layer that added 4 features (upscaling factor, 2 in our case, squared) that were used to reshape the data into a higher dimension. This feature makes pixel shuffling an adaptive up sampling method since it can tune or learn those up sampling features in the convolutional layer to best match the target dataset~\citep{Shi_2016}. This adaptive up sampling, which is known to work better than non-adaptive methods~\citep{Shi_2016}, is what we believe to be the reason pixel shuffling was able to remove artifacts and improve DroughtFormer performance over convolutional transpose or linear interpolation. Of additional note, the prediction performance was also examined for all start dates starting in MAMJJA and SONDJF separately, and while SONDJF months generally have better prediction skill, the difference between MAMJJA and SONDJF skill was always small (not shown). Thus the CrossFormer had a weak to non-existent seasonal dependence on predictive skill (in contrast to existing S2S forecast methods;~\citealt{Olaniyan_2025, Endris_2021, Phakula_2024}).

    Another one of the main contributions of DroughtFormer was its ability to predict climate anomalies (or deviations from the climatological mean). Prediction of climate anomalies are important for S2S forecasts in general (especially since climatology cannot give anomaly information), but especially in drought forecasting as it describes the relatively dryness or wetness of a location. The skill of DroughtFormer in predicting climate anomalies can be seen in the ACC (Fig.~\ref{fig:scorecard}), where DroughtFormer maintains consistently high scores of 0.85 or higher over Africa (in some cases 0.94+) for most variables. This suggests DroughtFormer was able to skillfully represented the deviations from climatology in space, and maintained that skill at long lead times. In general, DroughtFormer was found to be more responsive and dynamic to changes in the atmosphere than climatology predictions (found when comparing multiple DroughtFormer predictions to climatology predictions in the case study), allowing for good representation of climate anomalies.
    \par

\begin{figure}
        \noindent\fbox{\includegraphics[width=1.0\textwidth,angle=0]{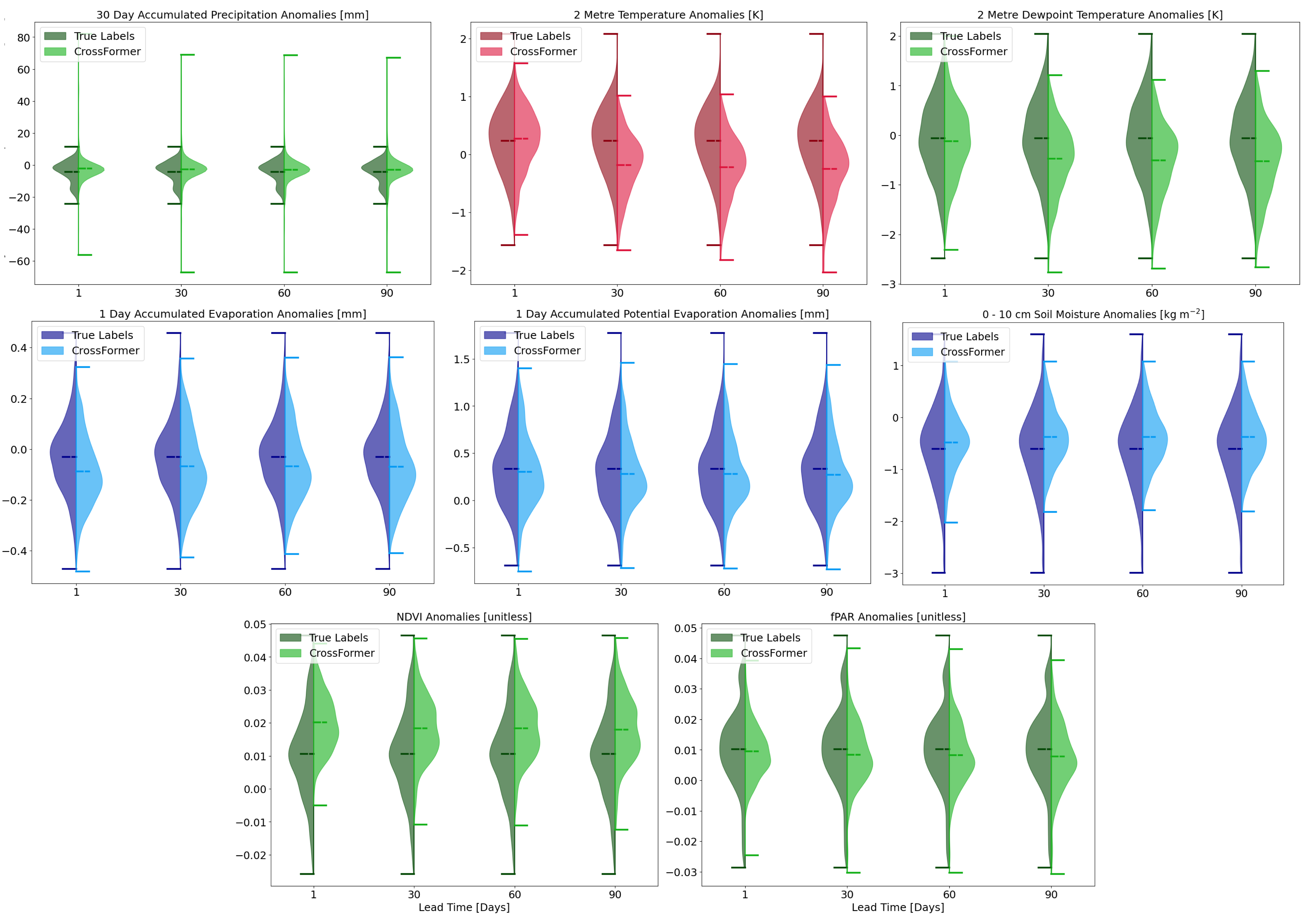}}\\
        \caption{Distribution of spatially averaged (over Africa) climate anomalies for multiple forecast lead times and from 2020 -- 2021 predictions. Variables include 30 day accumulated precipitation, temperature, dewpoint temperature, ET, PET, near surface SM (0 - 10 cm), NDVI, and fPAR. Dashed bars indicate the minimum, mean, and max of the distribution.}\label{fig:violin_fig}
    \end{figure}

    Further examination of the distribution of climate anomalies shows that, in general, DroughtFormer was able to accurately recreate and predict climate anomalies in time for near surface SM, fPAR, and ET (Fig.~\ref{fig:violin_fig}). There was, however, a positive bias in the mean distribution with NDVI and near surface SM and negative bias with ET, PET, fPAR, temperature, and dewpoint temperature. Additionally, DroughtFormer frequently struggled to capture the magnitude of the climate anomalies, or the distribution tails, often underestimating the anomalies magnitudes for ET, NDVI and near surface SM, and overestimating them for precipitation, while shifting the distribution down for PET, temperature, and dewpoint temperature. This implies our CrossFormer model was able to represent the occurrence of drought within a given region in Africa (as measured by SM, fPAR, ET, and PET), but struggled to capture the magnitude of the drought, often underestimating drought magnitudes or overestimating depending on the variable examined. This is particularly notable for precipitation where DroughtFormer struggled the most. This also has implications when forecasting FDs, as underestimating climate anomalies means DroughtFormer would fail to record the rapid development of drought conditions, and thus fail to detect FD when it occurs. While the converse is also possible for an over eager model, DroughtFormer was not found to be overly eager in predicting rapid drought development.

\begin{figure}
        \noindent\fbox{\includegraphics[width=1.0\textwidth,angle=0]{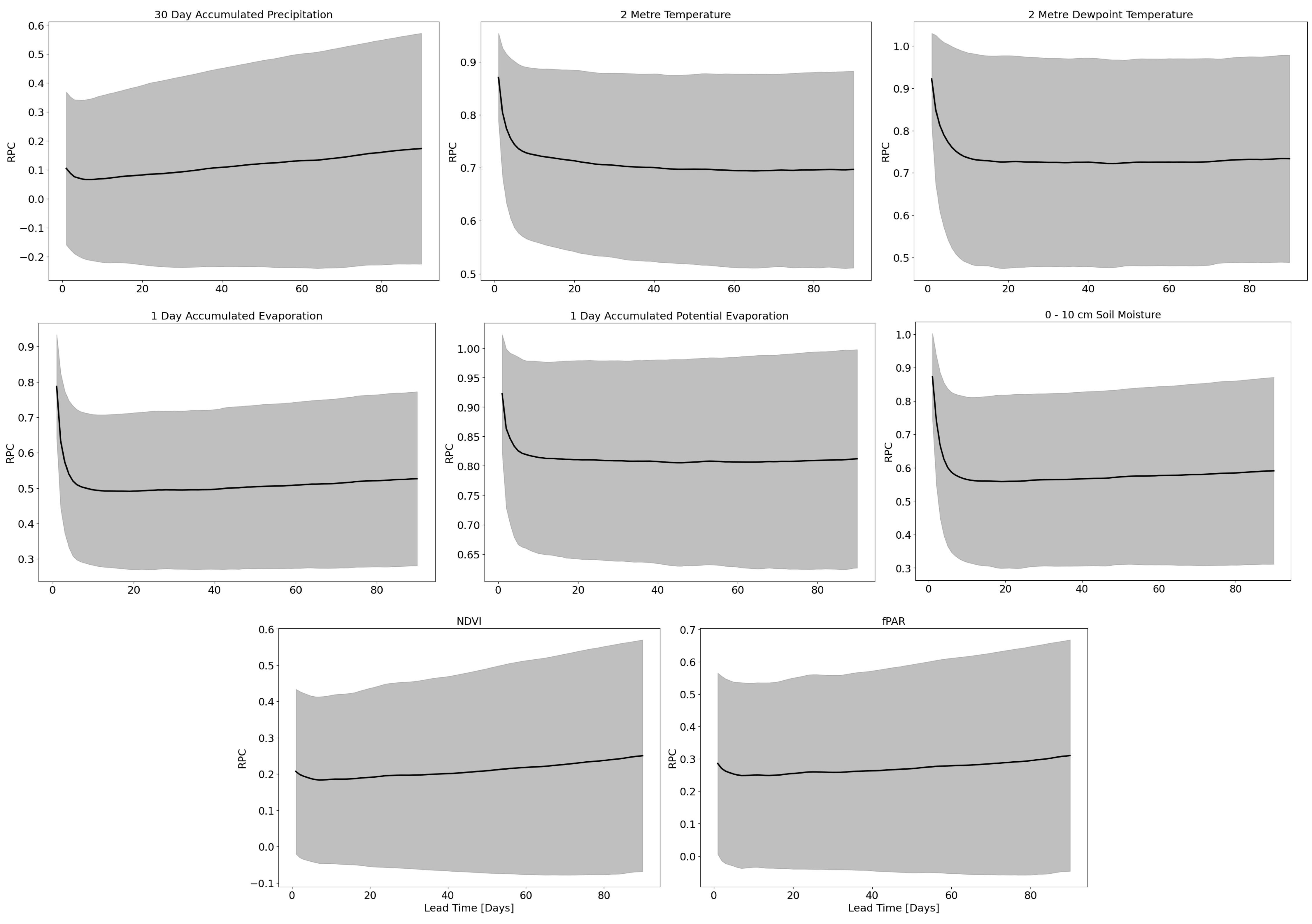}}\\
        \caption{Averaged (over Africa) RPC of climate anomalies for all forecasts from 2020 -- 2021 of 30 day accumulated precipitation, temperature, dewpoint temperature, ET, PET, near surface SM (0 - 10 cm), NDVI, and fPAR. Shading indicates 1 standard deviation in RPC values.}\label{fig:rpc_fig}
    \end{figure}

    This is further seen when quantitatively comparing the variation of climate anomalies to the true labels with the RPCs (Fig.~\ref{fig:rpc_fig}). For each variable, the RPC was below one, indicating DroughtFormer was overconfident in predicting the variation in climate anomalies~\citep{Scaife_2018, Brocker_2023}. While some predictions captured the anomaly variability well (e.g., temperature, dewpoint temperature, and PET), the majority of DroughtFormer predictions were under dispersive of the true climate anomaly variability, further emphasizing the limitations on capturing the magnitude of climate anomalies. This is more notable for climate anomalies, as the average RPC was found to be close one for most non-anomalous variables (not shown). 
    \par

\begin{figure}
        \noindent\fbox{\includegraphics[width=1.0\textwidth,angle=0]{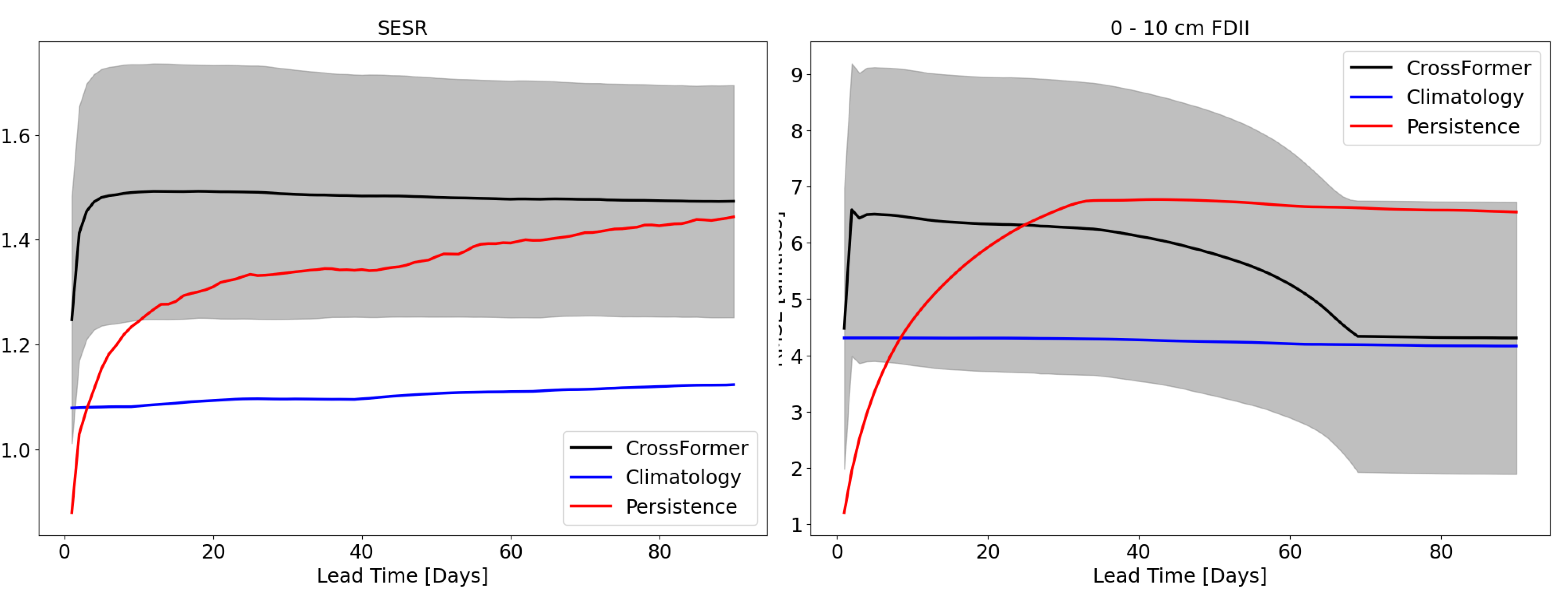}}\\
        \caption{Averaged RMSE over Africa for all forecasts from 2020 -- 2021 of SESR and FDII for the near surface (0 -- 10 cm) soil layer. Shading indicates 1 standard deviation in the CrossFormer predictive skill.}\label{fig:fd_rmse}
    \end{figure}

    Limitations in forecasting climate anomalies highlights one of the important ways DroughtFormer can be improved in future works as capturing the climate anomalies is especially important for drought and FD predictions. That is, we found errors in climate anomaly magnitudes would translate errors when examining FD indices (Fig.~\ref{fig:fd_rmse}). For Fig.~\ref{fig:fd_rmse} we used DroughtFormer's forecasts of ET, PET, and SM to forecast SESR and FDII. We found that, since the climate anomaly magnitudes were underestimated, DroughtFormer struggled to forecast the rapid drop off of moisture conditions necessary to predict FD. Additionally, DroughtFormer may fail to register a strong enough rapid intensification of drought conditions if its forecasts were initialized during that rapid intensification, yielding errors in 2 -- 30 day lead times (depending on when the rapid intensification and DroughtFormer forecast initializes). Errors in ET and PET also propagated into errors in SESR, resulting in the case where DroughtFormer struggled to beat persistence forecasts. Overall, this gives the three main limitations of DroughtFormer as: (1) It struggled to predict precipitation accumulations, (2) forecasting the magnitudes of climate anomalies, which results in (3) limitations in forecasting the occurrence of FD and FD indices, due to compounding errors in FD variables, and struggling to capture rapid intensification of drought conditions. A caveat to note is that while DroughtFormer's skill scores for SESR were not better than persistence, its case study forecasts were still useful in that DroughtFormer captured the location, timing, and sign of SESR (Fig.~\ref{fig:case_study_anomaly_sesr}).

    For non-anomalous variables, DroughtFormer was able to outperform climatology for a number of FD related variables, such as PET and surface level SM (Fig.~\ref{fig:rmse_fig}; though DroughtFormer skill dropped with increasing soil depth, in part due to fewer soil samples at deeper layers). In addition, our CrossFormer model produced skillful forecasts of dewpoint temperature and temperature (though it struggled more with tropical temperatures). It also performed comparable to, though slightly worse than, climatology for vegetation indices (however, it should be noted that DroughtFormer was effectively interpolating between the 8-day resolution of the MODIS variables, accumulating more error for each day). However, DroughtFormer struggled to outperform climatology with some of the more complex moisture variables, such as evaporation and precipitation, even at the 30-day accumulation levels (Figs.~\ref{fig:scorecard} -~\ref{fig:rmse_fig}). In general, The skill of DroughtFormer predictions were competitive with climatology for most variables (that is, only slightly worse than to better than climatology). However, skill of DroughtFormer being frequently close to climatology can also suggest that DroughtFormer is not heavily biased towards one regime, and any wet or dry bias in DroughtFormer stems from possible biases in the reanalyses datasets rather than the ML model.

    \section{2021 Case Study}\label{sec:case_study}
\begin{figure}
        \noindent\fbox{\includegraphics[width=1.0\textwidth,angle=0]{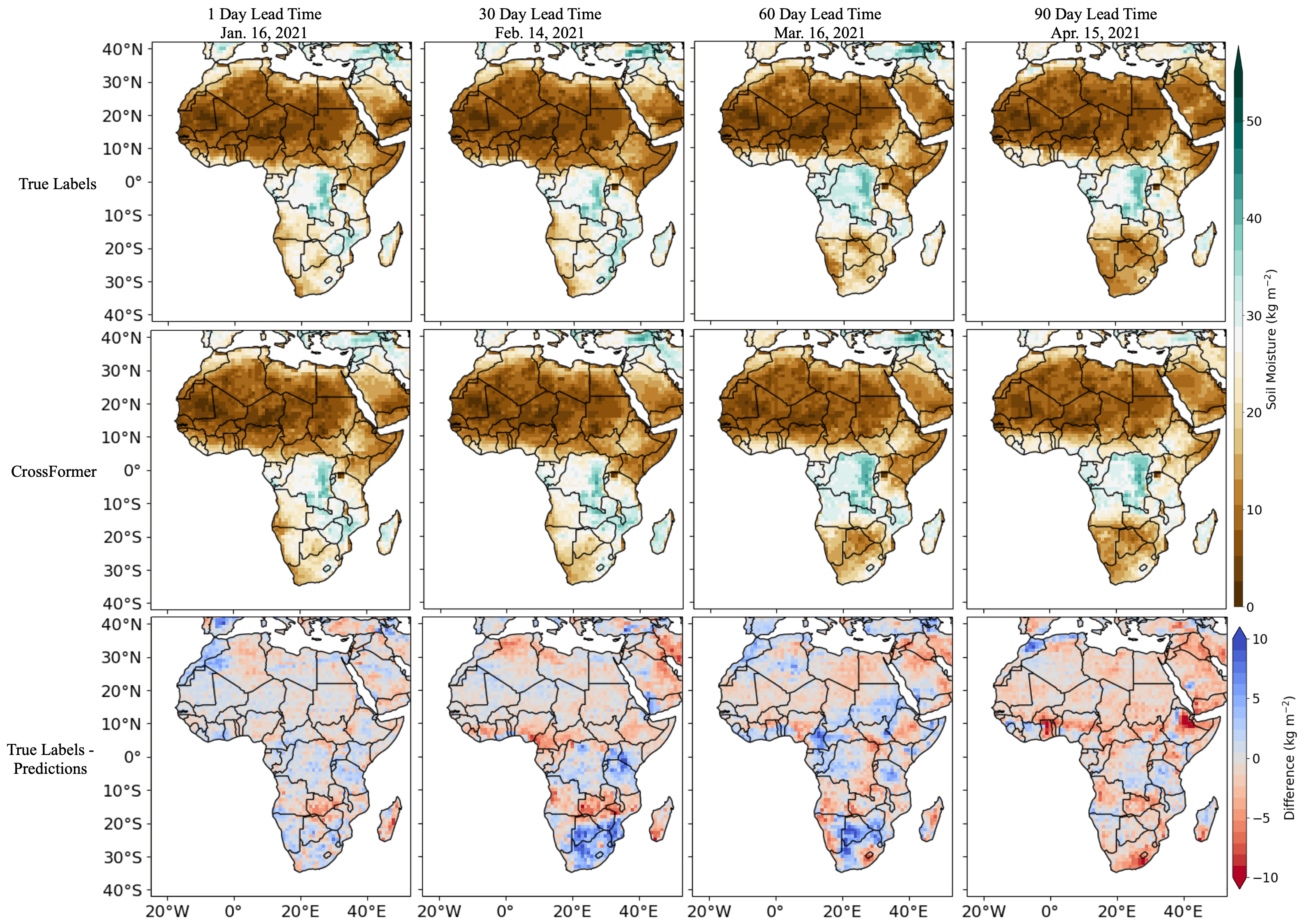}}\\
        \caption{True Labels (top row), CrossFormer predictions of (middle row), and difference between the two, (bottom row) for near surface soil moisture (0 - 10 cm) for 1, 30, 60, and 90 day lead times starting from January 15, 2021.}\label{fig:case_study}
    \end{figure}

\begin{figure}
        \noindent\fbox{\includegraphics[width=1.0\textwidth,angle=0]{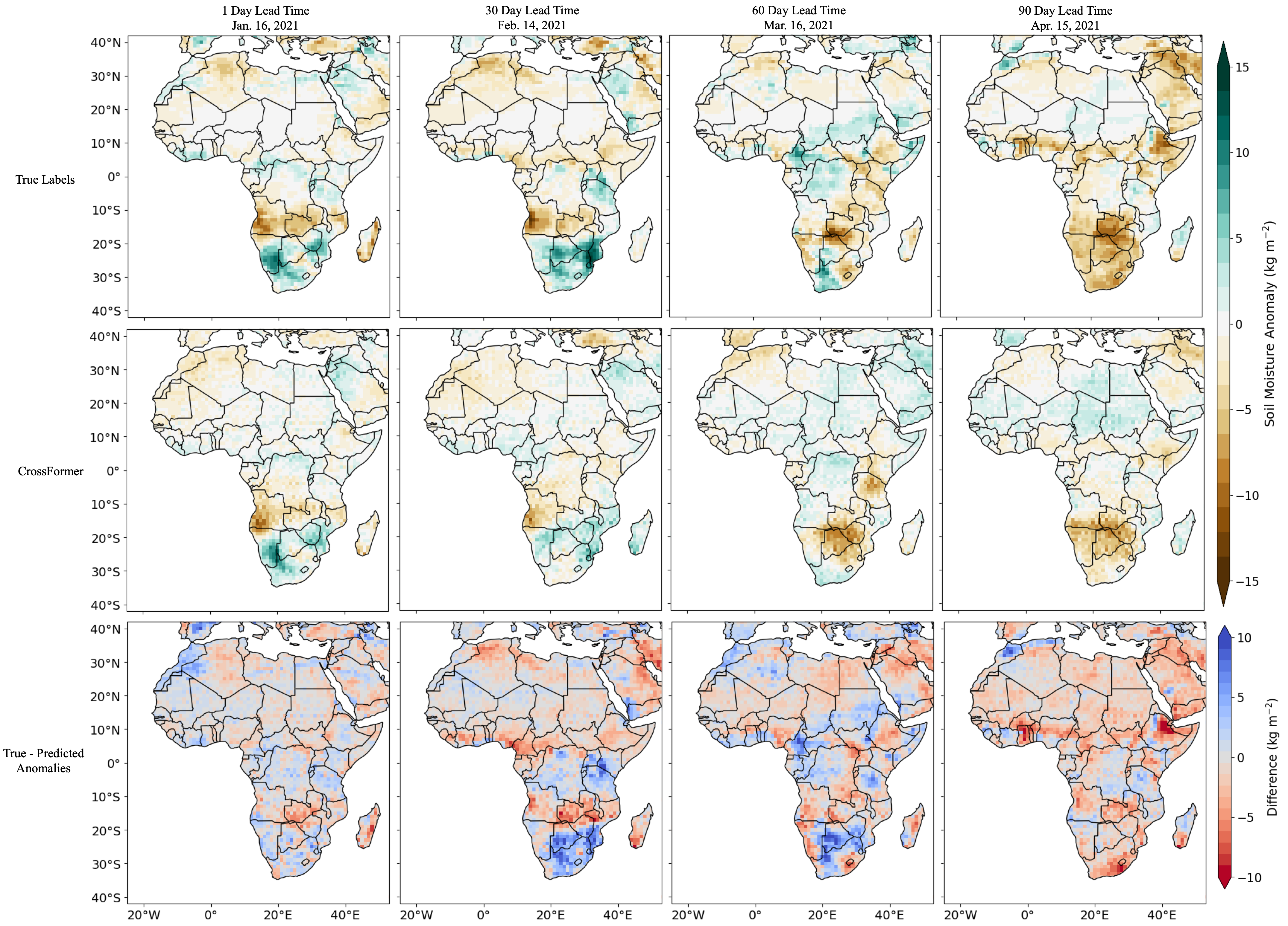}}\\
        \caption{True Labels (top row), CrossFormer predictions of (middle row), and difference between the two, (bottom row) for near surface soil moisture (0 - 10 cm) climate anomalies for 1, 30, 60, and 90 day lead times starting from January 15, 2021.}\label{fig:case_study_anomaly}
    \end{figure}

\begin{figure}
        \noindent\fbox{\includegraphics[width=1.0\textwidth,angle=0]{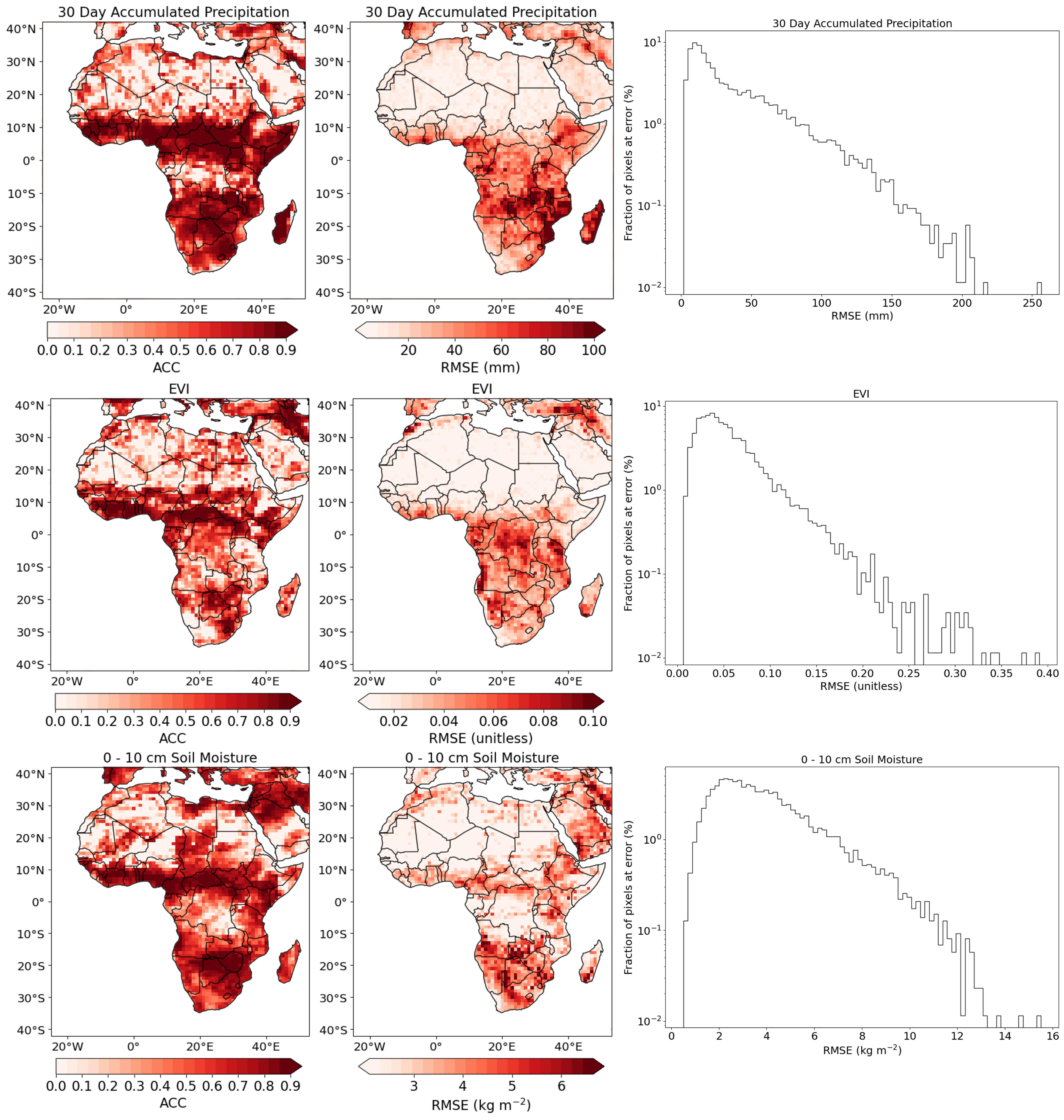}}\\
        \caption{Spatial Distribution (left) ACC and (middle) RMSE, averaged for all prediction times in the Jan. 15 - April 15 case study, for (top) 30 day precipitation accumulation, (middle) EVI, and (bottom) 0 - 10 cm soil moisture. (right) Percentage of RMSE distribution for each variable.}\label{fig:metric_in_space}
    \end{figure}

    In addition to the statistical results, a case study was also examined using January 2021 as the starting point for the model predictions. Figs.~\ref{fig:case_study} and~\ref{fig:case_study_anomaly} show the evolution of SM and SM anomalies, according to DroughtFormer predictions and true labels, and Fig.~\ref{fig:metric_in_space} shows the average skill across all prediction steps for the case study. Fig.~\ref{fig:case_study_anomaly} show the response of the soil moisture to the failure of the long rains that would normally precipitate in eastern Africa. The failure of the long rains was an extension of a historical drought that impacted millions of people in eastern Africa and left much the region without sufficient food for much of its population~\citep{Kimutai_2025, undrr_2024}. DroughtFormer accurately predicted the SM response indicating the presence of drought, and worsening of drought conditions when they are present, though it did have issues capturing the severity of the event. In addition, while Droughtformer failed to capture the precipitation anomalies for this event (not shown), it did capture SM deficits (Fig.~\ref{fig:case_study_anomaly}) and the vegetative response\footnote{While the vegetation index DroughtFormer showed the best predictive skill for was fPAR, EVI is shown here for case studies for illustrative purposes as it had a clearer signal.} to the lack of March and April rains (Fig.~\ref{fig:case_study_anomaly_evi_spring}) with which one could anticipate crop failure without drought mitigation strategies up to over 60 days in advance. This is of particular note, as DroughtFormer accurately predicting the sign, and general location and timing of negative anomalies is vital for identifying drought. This is further shown in Fig.~\ref{fig:metric_in_space}, where DroughtFormer error was small for SM and EVI for the case study (that is, high ACC and RMSE clustered around smaller values), showing good predictability for the event. DroughtFormer was also able to capture soil moisture and EVI deficits when in response to the failure of the short rains, initiating the drought, in late October of 2020 (Figs.~\ref{fig:case_study_anomaly_sm_fall} -~\ref{fig:case_study_anomaly_evi_fall}) over 67 days prior to the deficit, further showing DroughtFormer's usefulness in delivering early warning for this major drought event. 

    \par

    In addition, DroughtFormer was also able capture a drought that was present in southern Africa in the fall of 2020 (Figs.~\ref{fig:case_study_anomaly_sm_fall} -~\ref{fig:case_study_anomaly_evi_fall}) and spring of 2021 (Figs.~\ref{fig:case_study} -~\ref{fig:case_study_anomaly} and~\ref{fig:case_study_anomaly_evi_spring}) in response to a continuing rainfall shortage from 2018 and 2019. Similar to the drought in eastern Africa, DroughtFormer detected the deficit in soil moisture over 60 days in advance, indicating further rainfall shortage in October and again in Spring (though there was no signal in EVI degradation in Fall or Spring, suggesting a limited vegetation response in southern Africa).

    \section{Summary and Conclusions}\label{sec:conclusion}
    In summary, we have developed a state-of-the-art ML model, a modified CrossFormer, named DroughtFormer, for the purpose of drought prediction. In general, DroughtFormer showed impressive skill in representing drought related variables, with stable and skillful predictions out to 90 day lead times, with little dependence on month for skillful predictions. The model's predictive skill yields one of the main contributions of this study. The model's ability to stably predict multiple variables for the full S2S timescale is largely attributed to the CrossFormer model's up sampling method (pixel shuffling in this case), applied physics constraints, and the time scale explored (daily). That is, the data examined at the daily timescale (daily averaged or accumulations) presented an easier task for DroughtFormer to predict, yielding longer stable lead times. Future studies are encouraged to explore other time scales, as we hypothesize the use of pentads or weekly averages, often used in drought and FD monitoring, could potentially yield stable and skillful predictions for even longer lead times. In addition, DroughtFormer provided skillful predictions of multiple variables over Africa, and interpolation of vegetation indices, providing a baseline of deterministic forecasts of prediction skill for future studies to improve upon (Table~\ref{tab:baselines}). In some cases the model struggled to outperform climatology (e.g., evaporation, precipitation, and soil moisture for deeper layers), but still provided comparable and skillful forecasts for those variables (except for precipitation) with which climate anomalies could, importantly, be predicted. The caveat here is that DroughtFormer had difficulty in predicting the magnitudes of climates. We also found that increasing the number of static grid points made it harder for DroughtFormer to learn and predict the non-static values (e.g., DroughtFormer's skill dropped off in deeper soil layers, where there are fewer grid points with valid soil moisture values). This feature could be overcome with alternate neural networks that allow for heterogeneous grid structures, such as graph neural networks, that would remove static grid points. 
    \par

    Additionally, DroughtFormer was able to forecast the timing, location, and sign of climate anomalies within most forecasted variables (precipitation excepting), indicating the presence of drought weeks in advance. DroughtFormer used this in a case study for 2020 and 2021 to identify the presence of negative soil moisture anomalies, thus detecting drought, in response to a lack of precipitation over 60 days before they occurred. However, DroughtFormer struggled with the magnitude of the climate anomalies, often underestimating the magnitudes. This resulted in the three of the main limitations of DroughtFormer; it struggled in predicting precipitation accumulations, and the magnitudes in the climate anomalies. Errors in climate anomaly magnitudes also translated into errors in predicting rapid evolution of drought conditions, which limits its ability to detect and forecast FD. This is further compounded as errors in evaporation propagated into errors in evaporative stress, yielding the model's difficulty in predicting SESR, and any FD related to evaporative stress. Hence, DroughtFormer was able to accurately represent drought events, but was more limited in representing FD. Other limitations in DroughtFormer include errors in the reanalysis models and satellite data, as errors and biases in these datasets would have propagated into DroughtFormer (especially since only one dataset was used per variable). Future works are encouraged to examine the ability of ML models to represent the climate anomalies, and thence allow for the detection of FD events on the S2S time scale. Additional future work can look to add additional model runs to DroughtFormer, creating an ensemble approach, or creating an ensemble based ML model (as was recently added to the CREDIT framework;~\citealt{Schreck_2025}), which can also deliver probabilistic forecasts for more accurate S2S results. A multi-model approach would also allow an opportunity to explore the predictability of FD, or to tune the ensemble to better represent and focus on FD. These additions can help DroughtFormer create more accurate forecasts of drought and FD and thus help deliver forecasts that, when combined with working with local forecast offices, decision-makers, end-users, and weather-sensitive sectors, can help mitigate the impact of the severe droughts that occur within Africa.

    \acknowledgments
    This material is based upon work supported by the U.S. National Science Foundation under Grant No. RISE-2019758 and the Google Foundation. Any opinions, findings, and conclusions or recommendations expressed in this material are those of the author(s) and do not necessarily reflect the views of the U.S. National Science Foundation.
    \par
    
    Our thanks go to the CREDIT team for developing the CREDIT toolkit, and to Dr. Gagne and Dr. Schreck for helping us to get CREDIT models running and their assistance in debugging other issues.
    \par

    The computing for this project was performed at the OU Supercomputing Center for Education \& Research (OSCER) at the University of Oklahoma (OU).

    \datastatement
    ERA5 data for this project can be collected from the Copernicus data store~\url{https://cds.climate.copernicus.eu/datasets} or from the Google Cloud public datasets~\url{https://docs.cloud.google.com/storage/docs/public-datasets}. GLDAS2 and IMERG data can be found at NASA's EarthData repository~\url{https://disc.gsfc.nasa.gov/datasets/GLDAS_NOAH025_3H_2.1/summary} and~\url{https://disc.gsfc.nasa.gov/datasets/GPM_3IMERGDF_07/summary} respectively. MODIS data can be downloaded from EarthData at~\url{www.earthdata.nasa.gov/data/catalog/lpcloud-mod09a1-061} for reflectance observations and~\url{https://www.earthdata.nasa.gov/data/catalog/laads-mod15a2gfs-6} for fPAR and LAI observations. Lastly, climate indices were collected from the NOAA's Physical Sciences Laboratory datasets, with ENSO data collected from~\url{https://psl.noaa.gov/data/climateindices/list/} and IOD from~\url{https://psl.noaa.gov/data/timeseries/month/DMI/}. Code used to collect and process the data, and post process model forecasts is freely available at~\url{https://github.com/Rarell/global-drought-prediction-with-credit}. 

    \appendix
    \renewcommand{\thesection}{\Alph{section}}
    \renewcommand{\thefigure}{\thesection\arabic{figure}}
    \section{CrossFormer Parameter Tuning}\label{sec:appendix_a}
    Parameter tuning for the CrossFormer was performed using a smaller form of the model. That is, testing was done by training the model for 50 epochs of training on single-step training on 10 years of data (2005 -- 2015) and 2 years of validation data (2016 -- 2017), while 2018 was treated as the test set. The results of the test set were compared to the test set performance of the best performing model at that point. Within this context, multiple different model parameters were examined and tested to try and better tune the model. This included the modification of the model architecture (adding and removing attention heads), up sampling methods, physics constraints, single-step and multistep training, different loss functions and modifying latitude weights to focus on Africa, use of higher spatial resolution data, and use of additional climate indices.
    \par

\begin{figure}[!htbp]
        \noindent\fbox{\includegraphics[width=1.0\textwidth,angle=0]{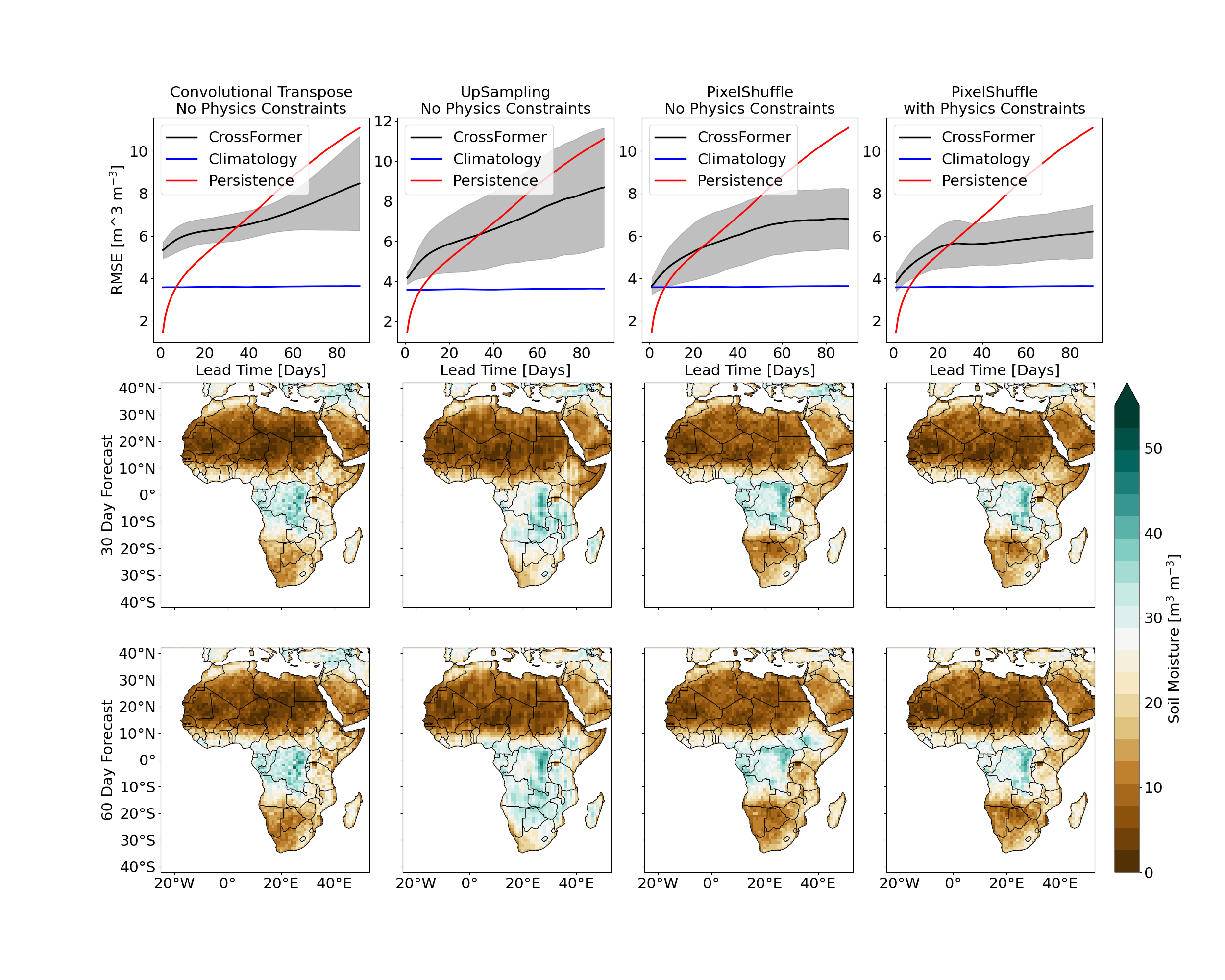}}\\
        \caption{Predictive skill of near surface (0 -- 10 cm) soil moisture for the CrossFormer with different up sampling methods and with and without physics constraints. (far left column) Using convolutional transpose up sampling and no physics, (second left column) UpSampling up sampling and no physics, (second right column) pixel shuffling up sampling and no physics, and (far right column) pixel shuffling up sampling with physics constraints. (top row) Average RMSE skill as function of lead time and compared to persistence and climatology, (middle row) 30 day forecast of near surface soil moisture starting from January 15, 2021, and (bottom row) 60 day forecast of near surface soil moisture. Note for the purposes of illustration, these results are from CrossFormer models trained on 5 years of data each.}\label{fig:a1}
    \end{figure}

    The most notable impact to model performance was found to be the up sampling method and use of physics constraints. At the start of the project, the CrossFormer model was limited in its prediction ability by the development of unphysical, artifacts that would overtake the predictions at 30+ day lead times. Interestingly, these artifacts appeared as striping artifacts when predicting ERA5 variables, and random noise injection when predicting GLDAS2 variables (e.g., left most column in Fig.~\ref{fig:a1}). We tried using UpSampling layers then, as the linear interpolation has generally been shown to do better than convolutional transpose. However, the artifacts remained with UpSampling layers, in the form of vertical striping, in predictions of ERA5 and GLDAS2 variables (Fig.~\ref{fig:a1}, second left column). Following correspondence with the CREDIT team, pixel shuffling was incorporated as the up sampling method, which successfully removed the striping artifacts within the model as well as greatly improved model accuracy (Fig.~\ref{fig:a1}, second right column). In addition, the use of physics constraints was also implemented as they were shown to improve stability in longer lead times~\citep{Watt-Meyer_2024, Sha_2025}. This was reflected in our CrossFormer model as well, where the physics constraints improved model stability to the point it was able to make stable and accurate 1 day predictions for over 90 days, with the lowest accuracy score occurring around 20 day lead times (demonstrating strong stability while maintaining realistic predictions; Fig.~\ref{fig:a1}, far right column). 
    \par

    Multistep training was also explored to try to increase model forecast skill above climatology. After testing several multistep forecast lengths, we found multistep training was found to greatly improve forecast skill over just using single-step training. The length of the single-step pre-training had minimal impact as the multistep training quickly converged to its solution. However, increasing the number of steps had two additional side effects: (1) It increased the training time by a factor of the number of steps, and (2) increasing the number of steps also increased the error for each step added (which could prematurely trigger early stopping). For a large number of steps, the gradient accumulation was large enough to cause the model to diverge (our CrossFormer exhibited this behavior when more than 5 steps were used). In addition to these side effects, it was found that adding additional steps had minimal impact on the final results. Thus, the simpler 3 step multistep training was chosen for our model. 
    \par

    Several other parameters were explored, however they were not implemented in the final model as they did not improve test results, had a detrimental effect, or had mixed results (i.e., small improvements in some variables, and small detrimental effects in others). For example, we explored the MAE and Huber loss functions in the hopes of improving predictions of outlier data points (i.e., extremes). However, there was not any significant improvement over the MSE loss function. Additionally, reducing latitude weights outside of Africa was also experimented with to help the model focus on the desired domain. However, these experiments failed to yield any positive or significant improvements over the African domain.
    \par

    The standard resolution if the ERA5 model ($0.25^{\circ} \times 0.25^{\circ}$) was also explored, but the CrossFormer had a difficult time learning the finer grain details while the higher resolution increased computation time by more than order of magnitude. In addition, several other concessions had to be made to allow the higher resolution model to fit on available RAM space (such as reducing the number of attention heads by half), which could have negatively impacted the higher resolution results as well. Lastly, other climate indices were also explored to improve model predictions. Namely the PDO, NAO, and AMO were explored, but the addition of extra variables that were static for $\sim$30 time steps for each grid point resulted in deteriorating model performance. 
    Thus, these additional indices were excluded from the final model run.

    \section{DroughtFormer Training Details}\label{sec:appendix_b}
    \renewcommand{\thefigure}{\thesection\arabic{figure}}
    \setcounter{figure}{0}
    \renewcommand{\thetable}{\thesection\arabic{table}}
    \setcounter{table}{0}
    Both single and multistep training steps were trained with cosine-annealing learning rates, with a maximum learning rate of $10^{-4}$ and a latitude weighted mean squared error (MSE) loss function. The model was also trained with the AdamW optimizer with batch sizes of 4. 

    \begin{table}[]
        \hspace{-0.0cm}
        \caption{Model details regarding training, regularization, and computation parameters and specifications for the CrossFormer model.}
        \small
        \centering
        \begin{tabular}{|R{4.0cm}|C{2.3cm}C{2.3cm}|}
            \hline
            & Singlestep Training & Multistep Training \\
            \hline
            Epochs & 120 & 100 \\
            Rollout & 1 & 3 \\
            GPU Memory per Model Instance & $\sim$28.1 GB & $\sim$28.1 GB \\
            Training Time (Hours) & 68 & 224 \\ 
            Total GPU Hours & 272 & 896 \\
            \hline
            & \multicolumn{2}{|c|}{Single Step and Multistep Training} \\
            \hline
            Learining Rate ($\times$10$^{-4}$) &    \multicolumn{2}{|l|}{\hspace{2.3cm} 1} \\
            L2 Regularization ($\times$10$^{-5}$) & \multicolumn{2}{|l|}{\hspace{2.3cm} 1} \\
            Attention \& Feed Forward Dropout & \multicolumn{2}{|l|}{\hspace{2.3cm} 0.05} \\
            Batch Size & \multicolumn{2}{|l|}{\hspace{2.3cm} 4} \\
            Number of Parameters & \multicolumn{2}{|l|}{\hspace{2.3cm} 179M} \\
            GPU Nodes & \multicolumn{2}{|l|}{\hspace{2.3cm} 4} \\
            \hline
        \end{tabular}
        \label{tab:training_parameters}
    \end{table}

    \newpage

    \section{Baseline for Deterministic Forecasting in Africa}\label{sec:appendix_c}
    \renewcommand{\thefigure}{\thesection\arabic{figure}}
    \setcounter{figure}{0}
    \renewcommand{\thetable}{\thesection\arabic{table}}
    \setcounter{table}{0}

    Overall, DroughtFormer showed skillful forecasts in a number of non-anomalous variables with Table~\ref{tab:baselines} showing the RMSE and ACC scores for multiple variables at several lead times, yielding a baseline of deterministic forecasts for future works to build upon and beat. In addition, these same patterns and trends discussed in Section~\ref{sec:S2S_pred} were observed when examining the mean absolute error as well (not shown).

    \begin{table}[]
        \hspace{-0.0cm}
        \caption{Baseline RMSE (left of slash) and ACC (right of slash) scores for the drought related variables outputted by the CrossFormer for multiple lead times. RMSE/ACC scores in parentheses are climatology skill scores when the CrossFormer failed to outperform climatology. Bolded RMSE values indicate a statistically significant difference from the climatology RMSE at the 95\% confidence level, according to the Diebold-Mariano test.}
        \small
        \centering
        \begin{tabular}{|R{2.8cm}|C{1.4cm}C{1.4cm}C{1.4cm}C{1.4cm}C{1.4cm}C{1.4cm}|}
            \hline
            Variable (units) & 1 Day & 10 Day & 30 Day & 50 Day & 70 Day & 90 Day \\
            \hline
            Temperature (K) & \textbf{1.57}/0.96 & 1.99/0.94 & 2.08/0.94\newline (2.02/0.94) & 2.10/0.94\newline (2.02/0.94) & 2.10/0.94\newline (2.02/0.94) & 2.11/0.94\newline (2.02/0.94) \\
            Dewpoint\newline Temperature (K) & \textbf{1.95}/0.98 & \textbf{2.70}/0.96 & \textbf{2.74}/0.95 & \textbf{2.74}/0.95 & \textbf{2.73}/0.95 & \textbf{2.73}/0.95 \\
            1 Day Accumulated\newline Precipitation (mm) & \textbf{8.12}/0.23\newline (\textbf{6.66}/0.42) & \textbf{7.85}/0.27\newline (\textbf{6.64}/0.42) & \textbf{7.86}/0.27\newline (\textbf{6.64}/0.42) & \textbf{7.82}/0.27\newline (\textbf{6.61}/0.42) & \textbf{7.79}/0.28\newline (\textbf{6.58}/0.42) & \textbf{7.80}/0.28\newline (\textbf{6.59}/0.42) \\
            30 Day Accumulated\newline Precipitation (mm) & \textbf{54.9}/0.85 (\textbf{46.2/0.88}) & \textbf{55.1}/0.85 (\textbf{46.1}/0.88) & \textbf{53.9}/0.85 (\textbf{46.1}/0.88) & \textbf{52.6}/0.86 (\textbf{46.0}/0.89) & \textbf{51.3}/0.87 (\textbf{45.6}/0.89) & \textbf{50.5}/0.87 (\textbf{45.6}/0.89) \\
            Evaporation (mm) & \textbf{0.78}/0.92 & \textbf{0.96}/0.88\newline (\textbf{0.86}/0.90) & \textbf{0.95}/0.88\newline (\textbf{0.86}/0.90) & \textbf{0.95}/0.88\newline (\textbf{0.87}/0.90) & \textbf{0.95}/0.88\newline (\textbf{0.87}/0.90) & \textbf{0.94}/0.88\newline (\textbf{0.87}/0.90)\\
            Potential\newline Evaporation (mm) & \textbf{1.69}/0.91 & \textbf{1.87}/0.89 & \textbf{1.87}/0.89 & \textbf{1.87}/0.89 & \textbf{1.87}/0.89 & \textbf{1.87}/0.89 \\
            SM Layer 1 (kg m$^{-2}$) & \textbf{2.77}/0.97 & 3.52/0.94 & 3.51/0.94 & 3.48/0.94 & 3.44/0.94 & 3.40/0.95 \\
            SM Layer 2 (kg m$^{-2}$) & 9.58/0.83\newline (9.41/0.83) & 10.2/0.81\newline (9.45/0.83) & 10.1/0.82\newline (9.48/0.83) & 9.88/0.83\newline (9.47/0.84) & 9.64/0.84\newline (9.50/0.84) & 9.43/0.84 \\
            SM Layer 3 (kg m$^{-2}$) & \textbf{7.67}/0.88\newline (\textbf{5.30}/0.94) & \textbf{7.86}/0.87\newline (\textbf{5.35}/0.94) & \textbf{7.91}/0.87\newline (\textbf{5.45}/0.94) & \textbf{7.87}/0.87\newline (\textbf{5.53}/0.93) & \textbf{7.85}/0.87\newline (\textbf{5.65}/0.93) & \textbf{7.82}/0.87\newline (\textbf{5.76}/0.93) \\
            SM Layer 4 (kg m$^{-2}$) & \textbf{7.65}/0.98\newline (\textbf{2.99}/1.00) & \textbf{7.67}/0.98\newline (\textbf{2.98}/1.00) & \textbf{7.67}/0.98\newline (\textbf{2.96}/1.00) & \textbf{7.67}/0.98\newline (\textbf{2.93}/1.00) & \textbf{7.68}/0.98\newline (\textbf{2.91}/1.00) & \textbf{7.67}/0.98\newline (\textbf{2.87}/1.00) \\
            NDVI (unitless) & \textbf{0.11}/0.87\newline (\textbf{0.10}/0.89) & \textbf{0.11}/0.87\newline (\textbf{0.10}/0.89) & \textbf{0.11}/0.87\newline (\textbf{0.10}/0.89) & \textbf{0.11}/0.87\newline (\textbf{0.10}/0.89) & \textbf{0.10}/0.88\newline (\textbf{0.10}/0.89) & \textbf{0.10}/0.88\newline (\textbf{0.10}/0.90) \\
            EVI (unitless) & 0.04/0.96 & \textbf{0.04}/0.95 & 0.04/0.96 & 0.04/0.96 & 0.04/0.96 & 0.04/0.96 \\
            LAI (unitless) & \textbf{0.65}/0.90\newline (\textbf{0.61}/0.92) & \textbf{0.65}/0.90\newline (\textbf{0.61}/0.92) & \textbf{0.64}/0.90\newline (\textbf{0.61}/0.91) & 0.63/0.91\newline (0.60/0.91) & 0.62/0.91\newline (0.61/0.91) & 0.61/0.91\newline (0.61/0.91) \\
            fPAR (unitless) & \textbf{0.10}/0.89\newline (\textbf{0.09}/0.90) & \textbf{0.10}/0.89\newline (\textbf{0.09}/0.90) & \textbf{0.10}/0.89\newline (\textbf{0.09}/0.90) & 0.10/0.89\newline (0.09/0.90) & 0.10/0.90\newline (0.09/0.90) & 0.10/0.90\newline (0.09/0.90)\\
            \hline
        \end{tabular}
        \label{tab:baselines}
    \end{table}

    \section{Additional Case Study Figures}\label{sec:appendix_d}
    \renewcommand{\thefigure}{\thesection\arabic{figure}}
    \setcounter{figure}{0}


    \begin{figure}[H]
        \noindent\fbox{\includegraphics[width=1.0\textwidth,angle=0]{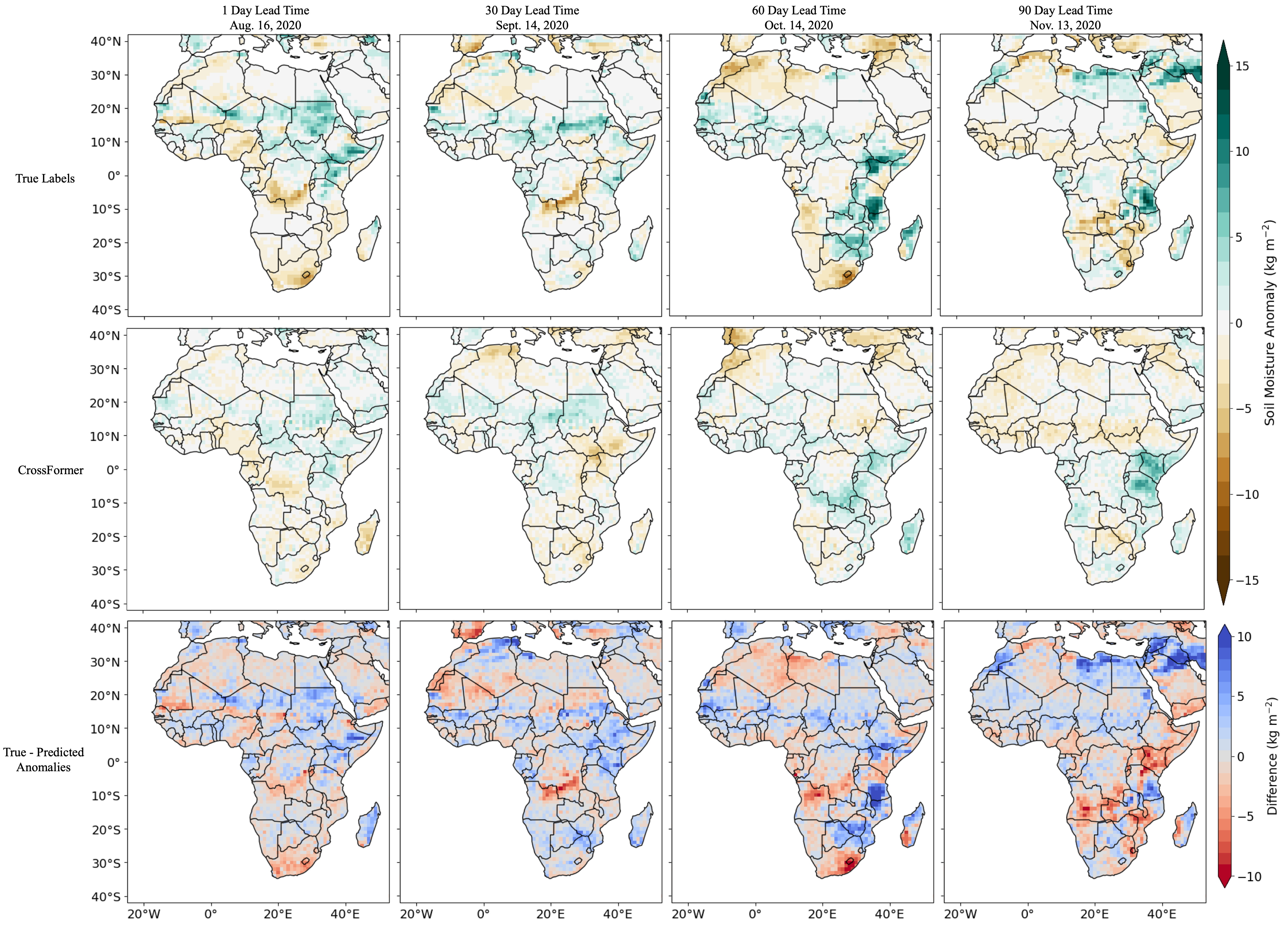}}\\
        \caption{True Labels (top row), CrossFormer predictions of (middle row), and difference between the two, (bottom row) for near surface soil moisture (0 - 10 cm) climate anomalies for 1, 30, 60, and 90 day lead times starting from August 15, 2020.}\label{fig:case_study_anomaly_sm_fall}
    \end{figure}


    \begin{figure}[H]
        \noindent\fbox{\includegraphics[width=1.0\textwidth,angle=0]{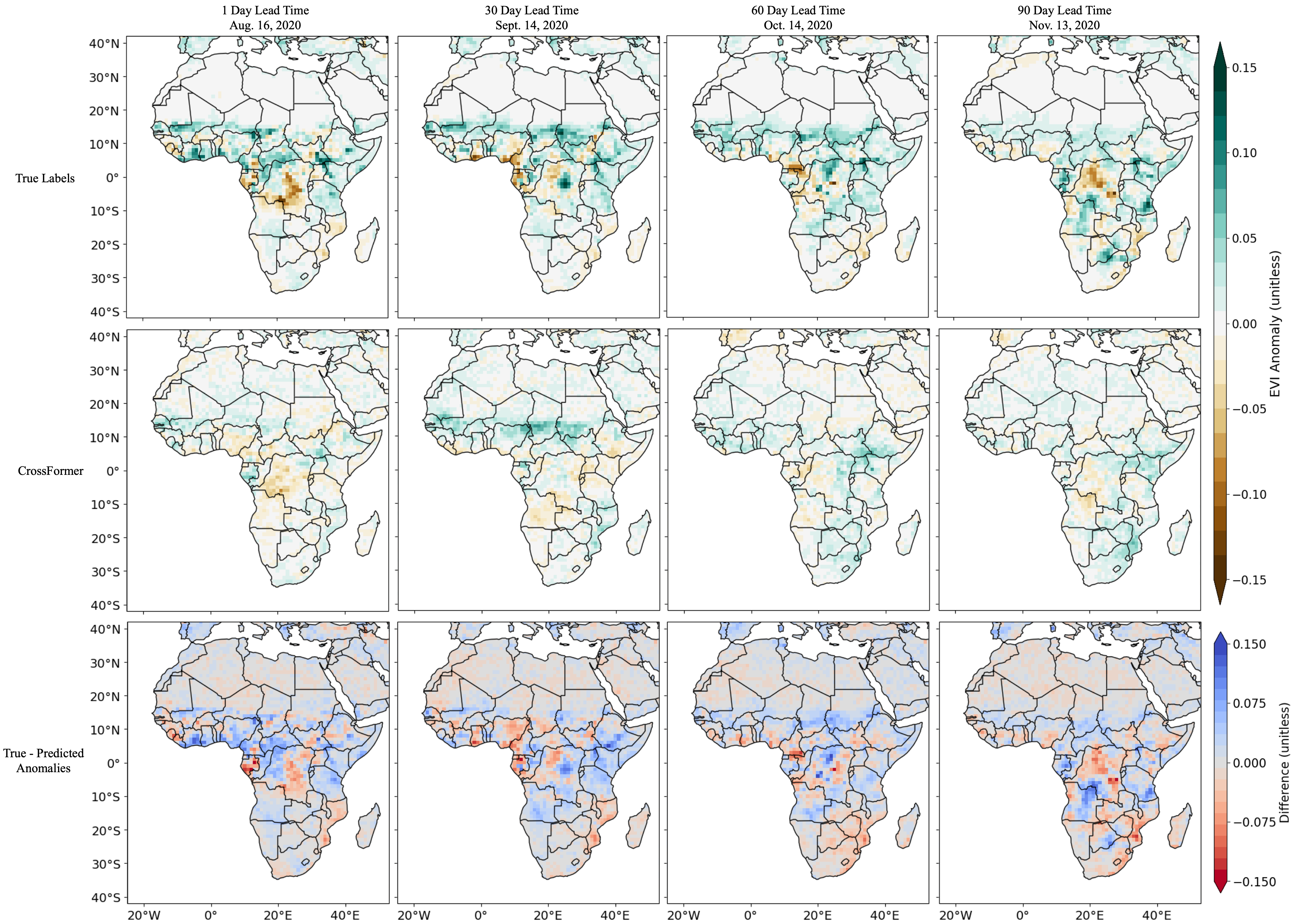}}\\
        \caption{True Labels (top row), CrossFormer predictions of (middle row), and difference between the two, (bottom row) for the Enhanced Vegetation Index climate anomalies for 1, 30, 60, and 90 day lead times starting from August 15, 2020.}\label{fig:case_study_anomaly_evi_fall}
    \end{figure}


    \begin{figure}[H]
        \noindent\fbox{\includegraphics[width=1.0\textwidth,angle=0]{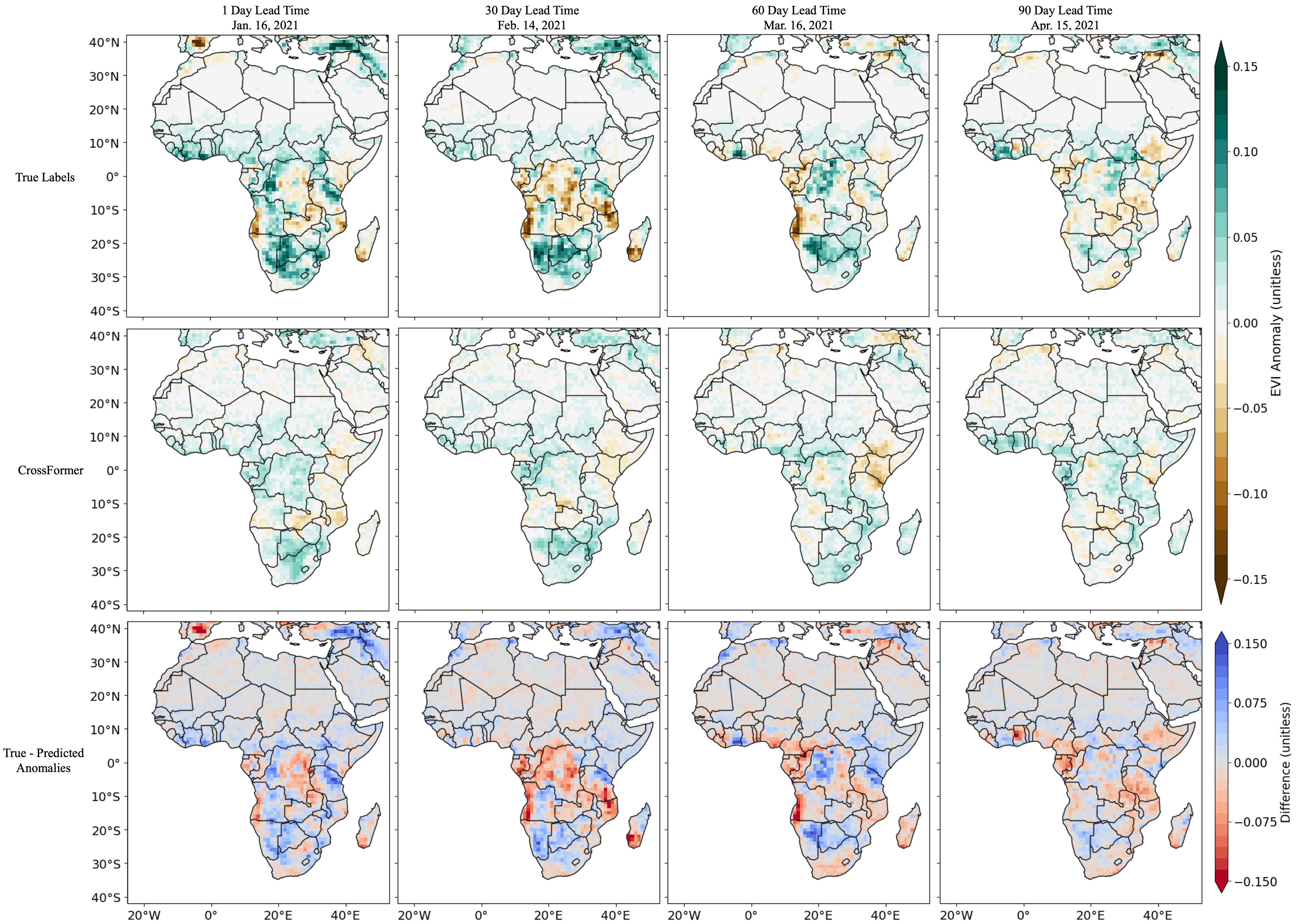}}\\
        \caption{True Labels (top row), CrossFormer predictions of (middle row), and difference between the two, (bottom row) for the Enhanced Vegetation Index climate anomalies for 1, 30, 60, and 90 day lead times starting from January 15, 2021.}\label{fig:case_study_anomaly_evi_spring}
    \end{figure}

    \begin{figure}[H]
        \noindent\fbox{\includegraphics[width=1.0\textwidth,angle=0]{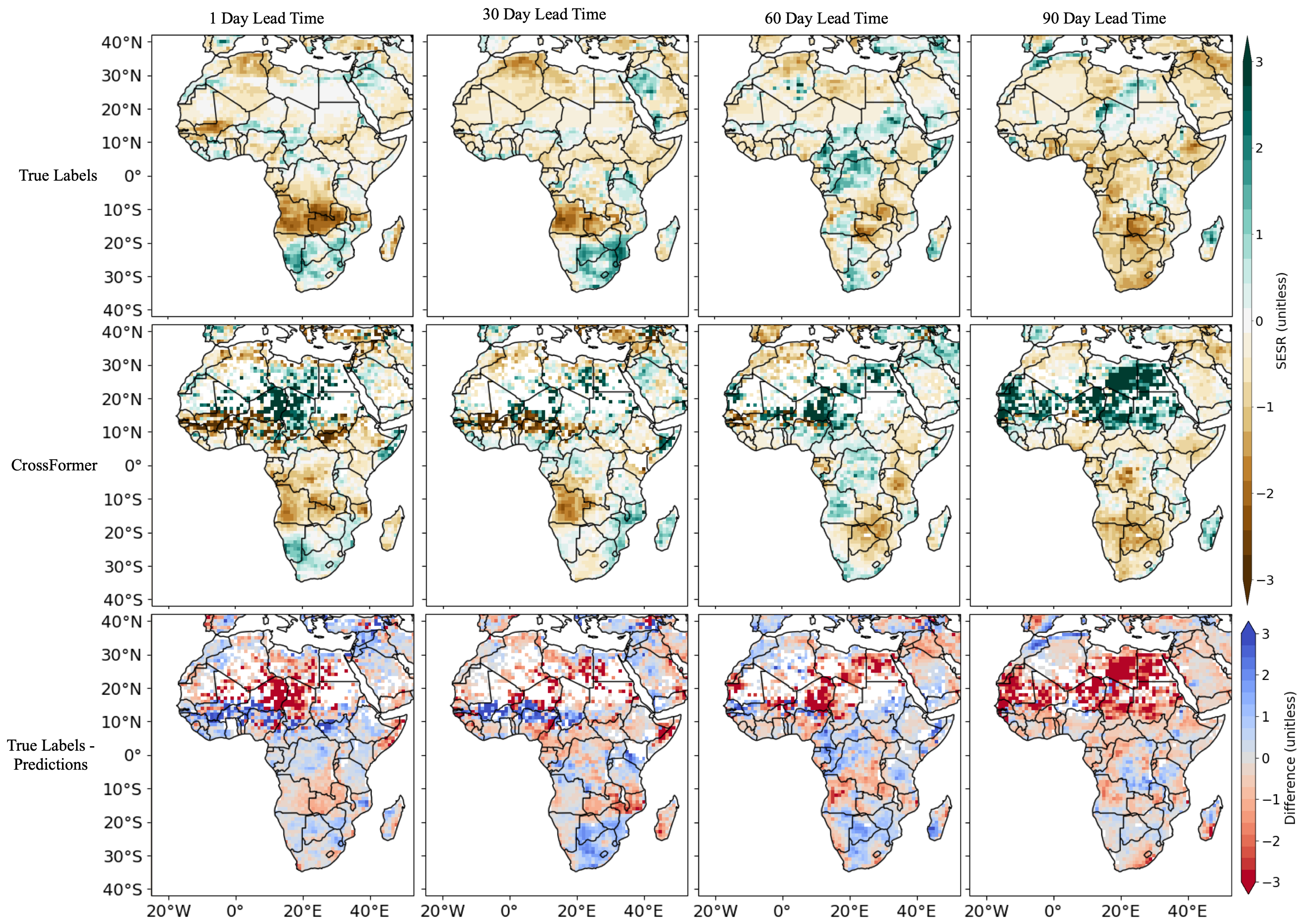}}\\
        \caption{True Labels (top row), CrossFormer predictions of (middle row), and difference between the two, (bottom row) for the Standardized Evaporative Stress Ratio for 1, 30, 60, and 90 day lead times starting from January 15, 2021.}\label{fig:case_study_anomaly_sesr}
    \end{figure}

    \bibliographystyle{ametsocV6}
    \bibliography{ML_Global_Drought}

\begin{thebibliography}{97}
\providecommand{\natexlab}[1]{#1}
\providecommand{\url}[1]{\texttt{#1}}
\renewcommand{\UrlFont}{\rmfamily}
\providecommand{\urlprefix}{URL }
\expandafter\ifx\csname urlstyle\endcsname\relax
  \providecommand{\doi}[1]{https://doi.org/\discretionary{}{}{}#1}\else
  \providecommand{\doi}{https://doi.org/\discretionary{}{}{}\begingroup \urlstyle{rm}\Url}\fi
\providecommand{\eprint}[2][]{\url{#2}}

\bibitem[{Adloff et~al.(2022)Adloff, Singer, MacLeod, Michaelides, Mehrnegar, Hansford, Funk,, and Mitchell}]{Adloff_2022}
Adloff, M., M.~B. Singer, D.~A. MacLeod, K.~Michaelides, N.~Mehrnegar, E.~Hansford, C.~Funk, and D.~Mitchell, 2022: Sustained water storage in horn of africa drylands dominated by seasonal rainfall extremes. \textit{Geophysical Research Letters}, \textbf{49~(21)}, \doi{10.1029/2022gl099299}.

\bibitem[{AghaKouchak et~al.(2022)}]{AghaKouchak_2022}
AghaKouchak, A., and Coauthors, 2022: Status and prospects for drought forecasting: opportunities in artificial intelligence and hybrid physical{\textendash}statistical forecasting. \textit{Philosophical Transactions of the Royal Society A: Mathematical, Physical and Engineering Sciences}, \textbf{380~(2238)}, \doi{10.1098/rsta.2021.0288}.

\bibitem[{Anande and Otkin(2026)Anande, and Otkin}]{Anande_2026}
Anande, D.~M., and J.~A. Otkin, 2026: Examining climatological characteristics and drivers of rapid change events in hydrological conditions over africa. \textit{106th Annual Meteorological Society Meeting}, 40th Conference on Hydrology.

\bibitem[{Ayugi et~al.(2022)}]{Ayugi_2022}
Ayugi, B., and Coauthors, 2022: Review of meteorological drought in africa: Historical trends, impacts, mitigation measures, and prospects. \textit{Pure and Applied Geophysics}, \textbf{179~(4)}, 1365--1386, \doi{10.1007/s00024-022-02988-z}.

\bibitem[{Basara et~al.(2019)Basara, Christian, Wakefield, Otkin, Hunt,, and Brown}]{Basara_2019}
Basara, J.~B., J.~I. Christian, R.~A. Wakefield, J.~A. Otkin, E.~H. Hunt, and D.~P. Brown, 2019: The evolution, propagation, and spread of flash drought in the central united states during 2012. \textit{Environmental Research Letters}, \textbf{14~(8)}, 084\,025, \doi{10.1088/1748-9326/ab2cc0}.

\bibitem[{Beaudoing et~al.(2020)Beaudoing, Rodell,, and {NASA/GSFC/HSL}}]{Beaudoing_2020_GLDAS}
Beaudoing, H., M.~Rodell, and {NASA/GSFC/HSL}, 2020: Gldas noah land surface model l4 3 hourly 0.25 x 0.25 degree, version 2.1. NASA Goddard Earth Sciences Data and Information Services Center, \urlprefix\url{https://disc.gsfc.nasa.gov/datacollection/GLDAS_NOAH025_3H_2.1.html}, \doi{10.5067/E7TYRXPJKWOQ}.

\bibitem[{Bi et~al.(2023)Bi, Xie, Zhang, Chen, Gu,, and Tian}]{Bi_2023}
Bi, K., L.~Xie, H.~Zhang, X.~Chen, X.~Gu, and Q.~Tian, 2023: Accurate medium-range global weather forecasting with 3d neural networks. \textit{Nature}, \textbf{619~(7970)}, 533--538, \doi{10.1038/s41586-023-06185-3}.

\bibitem[{Br{\"o}cker et~al.(2023)Br{\"o}cker, Charlton-Perez,, and Weisheimer}]{Brocker_2023}
Br{\"o}cker, J., A.~J. Charlton-Perez, and A.~Weisheimer, 2023: A statistical perspective on the signal‐to‐noise paradox. \textit{Quarterly Journal of the Royal Meteorological Society}, \textbf{149~(752)}, 911--923, \doi{10.1002/qj.4440}.

\bibitem[{Chapman et~al.(2025)Chapman, Schreck, Sha, Gagne, Kimpara, Zanna, Mayer,, and Berner}]{Chapman_2025}
Chapman, W.~E., J.~S. Schreck, Y.~Sha, D.~J. Gagne, D.~Kimpara, L.~Zanna, K.~J. Mayer, and J.~Berner, 2025: Camulator: Fast emulation of the community atmosphere model. arXiv, \urlprefix\url{https://arxiv.org/abs/2504.06007}, \doi{10.48550/ARXIV.2504.06007}.

\bibitem[{Chen et~al.(2023)Chen, Zhong, Zhang, Cheng, Xu, Qi,, and Li}]{Chen_2023}
Chen, L., X.~Zhong, F.~Zhang, Y.~Cheng, Y.~Xu, Y.~Qi, and H.~Li, 2023: Fuxi: a cascade machine learning forecasting system for 15-day global weather forecast. \textit{npj Climate and Atmospheric Science}, \textbf{6~(1)}, \doi{10.1038/s41612-023-00512-1}.

\bibitem[{Christian et~al.(2021)Christian, Basara, Hunt, Otkin, Furtado, Mishra, Xiao,, and Randall}]{Christian_2021}
Christian, J.~I., J.~B. Basara, E.~D. Hunt, J.~A. Otkin, J.~C. Furtado, V.~Mishra, X.~Xiao, and R.~M. Randall, 2021: Global distribution, trends, and drivers of flash drought occurrence. \textit{Nature Communications}, \textbf{12~(1)}, \doi{10.1038/s41467-021-26692-z}.

\bibitem[{Christian et~al.(2020)Christian, Basara, Hunt, Otkin,, and Xiao}]{Christian_2020}
Christian, J.~I., J.~B. Basara, E.~D. Hunt, J.~A. Otkin, and X.~Xiao, 2020: Flash drought development and cascading impacts associated with the 2010 russian heatwave. \textit{Environmental Research Letters}, \textbf{15~(9)}, 094\,078, \doi{10.1088/1748-9326/ab9faf}.

\bibitem[{Christian et~al.(2022)Christian, Basara, Lowman, Xiao, Mesheske,, and Zhou}]{Christian_2022}
Christian, J.~I., J.~B. Basara, L.~E. Lowman, X.~Xiao, D.~Mesheske, and Y.~Zhou, 2022: Flash drought identification from satellite-based land surface water index. \textit{Remote Sensing Applications: Society and Environment}, \textbf{26}, 100\,770, \doi{10.1016/j.rsase.2022.100770}.

\bibitem[{Christian et~al.(2019)Christian, Basara, Otkin, Hunt, Wakefield, Flanagan,, and Xiao}]{Christian_2019b}
Christian, J.~I., J.~B. Basara, J.~A. Otkin, E.~D. Hunt, R.~A. Wakefield, P.~X. Flanagan, and X.~Xiao, 2019: A methodology for flash drought identification: Application of flash drought frequency across the united states. \textit{Journal of Hydrometeorology}, \textbf{20~(5)}, 833--846, \doi{10.1175/jhm-d-18-0198.1}.

\bibitem[{Dale et~al.(2017)Dale, Fant, Strzepek, Lickley,, and Solomon}]{Dale_2017}
Dale, A., C.~Fant, K.~Strzepek, M.~Lickley, and S.~Solomon, 2017: Climate model uncertainty in impact assessments for agriculture: A multi‐ensemble case study on maize in sub‐saharan africa. \textit{Earth's Future}, \textbf{5~(3)}, 337--353, \doi{10.1002/2017ef000539}.

\bibitem[{Deman et~al.(2022)Deman, Koppa, Waegeman, MacLeod, Bliss~Singer,, and Miralles}]{Deman_2022}
Deman, V. M.~H., A.~Koppa, W.~Waegeman, D.~A. MacLeod, M.~Bliss~Singer, and D.~G. Miralles, 2022: Seasonal prediction of horn of africa long rains using machine learning: The pitfalls of preselecting correlated predictors. \textit{Frontiers in Water}, \textbf{4}, \doi{10.3389/frwa.2022.1053020}.

\bibitem[{Dikshit et~al.(2022{\natexlab{a}})Dikshit, Pradhan, Huete,, and Park}]{Dikshit_2022a}
Dikshit, A., B.~Pradhan, A.~Huete, and H.-J. Park, 2022{\natexlab{a}}: Spatial based drought assessment: Where are we heading? a review on the current status and future. \textit{Science of The Total Environment}, \textbf{844}, 157\,239, \doi{10.1016/j.scitotenv.2022.157239}.

\bibitem[{Dikshit et~al.(2022{\natexlab{b}})Dikshit, Pradhan,, and Santosh}]{Dikshit_2022b}
Dikshit, A., B.~Pradhan, and M.~Santosh, 2022{\natexlab{b}}: Artificial neural networks in drought prediction in the 21st century{\textendash}a scientometric analysis. \textit{Applied Soft Computing}, \textbf{114}, 108\,080, \doi{10.1016/j.asoc.2021.108080}.

\bibitem[{Doi et~al.(2022)Doi, Behera,, and Yamagata}]{Doi_2022}
Doi, T., S.~K. Behera, and T.~Yamagata, 2022: On the predictability of the extreme drought in east africa during the short rains season. \textit{Geophysical Research Letters}, \textbf{49~(22)}, \doi{10.1029/2022gl100905}.

\bibitem[{Dunstan et~al.(2025)}]{Dunstan_2025}
Dunstan, T., and Coauthors, 2025: Fastnet: Improving the physical consistency of machine-learning weather prediction models through loss function design. arXiv, \urlprefix\url{https://arxiv.org/abs/2509.17601}, \doi{10.48550/ARXIV.2509.17601}.

\bibitem[{Endris et~al.(2021)Endris, Hirons, Segele, Gudoshava, Woolnough,, and Artan}]{Endris_2021}
Endris, H.~S., L.~Hirons, Z.~T. Segele, M.~Gudoshava, S.~Woolnough, and G.~A. Artan, 2021: Evaluation of the skill of monthly precipitation forecasts from global prediction systems over the greater horn of africa. \textit{Weather and Forecasting}, \textbf{36~(4)}, 1275--1298, \doi{10.1175/waf-d-20-0177.1}.

\bibitem[{Fankhauser et~al.(2022)Fankhauser, Macharia, Coyle, Kathuni, McNally, Slinski,, and Thomas}]{Fankhauser_2022}
Fankhauser, K., D.~Macharia, J.~Coyle, S.~Kathuni, A.~McNally, K.~Slinski, and E.~Thomas, 2022: Estimating groundwater use and demand in arid kenya through assimilation of satellite data and in-situ sensors with machine learning toward drought early action. \textit{Science of The Total Environment}, \textbf{831}, 154\,453, \doi{10.1016/j.scitotenv.2022.154453}.

\bibitem[{Ferchichi et~al.(2024)Ferchichi, Chihaoui,, and Ferchichi}]{Ferchichi_2024}
Ferchichi, A., M.~Chihaoui, and A.~Ferchichi, 2024: Spatio-temporal modeling of climate change impacts on drought forecast using generative adversarial network: A case study in africa. \textit{Expert Systems with Applications}, \textbf{238}, 122\,211, \doi{10.1016/j.eswa.2023.122211}.

\bibitem[{Ford et~al.(2023)Ford, Otkin, Quiring, Lisonbee, Woloszyn, Wang,, and Zhong}]{Ford_2023}
Ford, T.~W., J.~A. Otkin, S.~M. Quiring, J.~Lisonbee, M.~Woloszyn, J.~Wang, and Y.~Zhong, 2023: Flash drought indicator intercomparison in the united states. \textit{Journal of Applied Meteorology and Climatology}, \textbf{62~(12)}, 1713--1730, \doi{10.1175/jamc-d-23-0081.1}.

\bibitem[{Foroumandi et~al.(2024)Foroumandi, Gavahi,, and Moradkhani}]{Foroumandi_2024}
Foroumandi, E., K.~Gavahi, and H.~Moradkhani, 2024: Generative adversarial network for real‐time flash drought monitoring: A deep learning study. \textit{Water Resources Research}, \textbf{60~(5)}, \doi{10.1029/2023wr035600}.

\bibitem[{Friedl and Sulla-Menashe(2022)Friedl, and Sulla-Menashe}]{Fruedl_2022_MODIS}
Friedl, M., and D.~Sulla-Menashe, 2022: Modis/terra+aqua land cover type yearly l3 global 0.05deg cmg v061. NASA EOSDIS Land Processes Distributed Active Archive Center, \urlprefix\url{https://lpdaac.usgs.gov/products/mcd12c1v061/}, \doi{10.5067/MODIS/MCD12C1.061}.

\bibitem[{Funk et~al.(2018)}]{Funk_2018}
Funk, C., and Coauthors, 2018: Examining the role of unusually warm indo‐pacific sea‐surface temperatures in recent african droughts. \textit{Quarterly Journal of the Royal Meteorological Society}, \textbf{144~(S1)}, 360--383, \doi{10.1002/qj.3266}.

\bibitem[{Gou et~al.(2022)}]{Gou_2022}
Gou, Q., and Coauthors, 2022: Application of an improved spatio-temporal identification method of flash droughts. \textit{Journal of Hydrology}, \textbf{604}, 127\,224, \doi{10.1016/j.jhydrol.2021.127224}.

\bibitem[{Guido et~al.(2020)}]{Guido_2020}
Guido, Z., and Coauthors, 2020: Farmer forecasts: Impacts of seasonal rainfall expectations on agricultural decision-making in sub-saharan africa. \textit{Climate Risk Management}, \textbf{30}, 100\,247, \doi{10.1016/j.crm.2020.100247}.

\bibitem[{Guo et~al.(2024)Guo, Zhou, Jiang, Wu, Sun,, and Jin}]{Guo_2024}
Guo, Y., T.~Zhou, W.~Jiang, B.~Wu, L.~Sun, and R.~Jin, 2024: Maximizing the impact of deep learning on subseasonal-to-seasonal climate forecasting: The essential role of optimization. arXiv, \urlprefix\url{https://arxiv.org/abs/2411.16728}, \doi{10.48550/ARXIV.2411.16728}.

\bibitem[{Hersbach et~al.(2023)}]{Hersbach_2023}
Hersbach, H., and Coauthors, 2023: Era5 hourly data on single levels from 1940 to present. Copernicus Climate Change Service (C3S) Climate Data Store (CDS), \urlprefix\url{https://cds.climate.copernicus.eu/doi/10.24381/cds.adbb2d47}, \doi{10.24381/CDS.ADBB2D47}.

\bibitem[{Hoell et~al.(2013)Hoell, Funk,, and Barlow}]{Hoell_2013}
Hoell, A., C.~Funk, and M.~Barlow, 2013: The regional forcing of northern hemisphere drought during recent warm tropical west pacific ocean la ni{\~n}a events. \textit{Climate Dynamics}, \textbf{42~(11--12)}, 3289--3311, \doi{10.1007/s00382-013-1799-4}.

\bibitem[{Holland et~al.(2023)Holland, Livneh,, and Thomas}]{Holland_2023}
Holland, M., B.~Livneh, and E.~Thomas, 2023: Performance of regression and artificial neural network models, informed with an in situ sensor network, in forecasting groundwater abstraction in the central valley, california. \textit{ACS ES\&T Water}, \textbf{3~(12)}, 3893--3904, \doi{10.1021/acsestwater.3c00322}.

\bibitem[{Huang et~al.(2017)}]{Huang_2017_ersst}
Huang, B., and Coauthors, 2017: Noaa extended reconstructed sea surface temperature (ersst), version 5. NOAA National Centers for Environmental Information, \urlprefix\url{https://www.ncei.noaa.gov/access/metadata/landing-page/bin/iso?id=gov.noaa.ncdc:C00927}, \doi{10.7289/V5T72FNM}.

\bibitem[{Ishii et~al.(2005)Ishii, Shouji, Sugimoto,, and Matsumoto}]{ishii_2005_cobe}
Ishii, M., A.~Shouji, S.~Sugimoto, and T.~Matsumoto, 2005: Objective analyses of sea‐surface temperature and marine meteorological variables for the 20th century using icoads and the kobe collection. \textit{International Journal of Climatology}, \textbf{25~(7)}, 865--879, \doi{10.1002/joc.1169}.

\bibitem[{Jiang et~al.(2021)Jiang, Zhou, Roundy, Hua,, and Raghavendra}]{Jiang_2021}
Jiang, Y., L.~Zhou, P.~E. Roundy, W.~Hua, and A.~Raghavendra, 2021: Increasing influence of indian ocean dipole on precipitation over central equatorial africa. \textit{Geophysical Research Letters}, \textbf{48~(8)}, \doi{10.1029/2020gl092370}.

\bibitem[{Kew et~al.(2021)}]{Kew_2021}
Kew, S.~F., and Coauthors, 2021: Impact of precipitation and increasing temperatures on drought trends in eastern africa. \textit{Earth System Dynamics}, \textbf{12~(1)}, 17--35, \doi{10.5194/esd-12-17-2021}.

\bibitem[{Kimutai et~al.(2025)}]{Kimutai_2025}
Kimutai, J., and Coauthors, 2025: Human-induced climate change increased 2021--2022 drought severity in horn of africa. \textit{Weather and Climate Extremes}, \textbf{47}, 100\,745, \doi{10.1016/j.wace.2025.100745}.

\bibitem[{Lang et~al.(2024{\natexlab{a}})}]{Lam_2024}
Lang, S., and Coauthors, 2024{\natexlab{a}}: Aifs -- ecmwf's data-driven forecasting system. arXiv, \urlprefix\url{https://arxiv.org/abs/2406.01465}, \doi{10.48550/ARXIV.2406.01465}.

\bibitem[{Lang et~al.(2024{\natexlab{b}})}]{Lang_2024}
Lang, S., and Coauthors, 2024{\natexlab{b}}: Aifs -- ecmwf's data-driven forecasting system. arXiv, \urlprefix\url{https://arxiv.org/abs/2406.01465}, \doi{10.48550/ARXIV.2406.01465}.

\bibitem[{Liebmann et~al.(2012)Liebmann, Blad{\'e}, Kiladis, Carvalho, B.~Senay, Allured, Leroux,, and Funk}]{Liebmann_2012}
Liebmann, B., I.~Blad{\'e}, G.~N. Kiladis, L.~M.~V. Carvalho, G.~B.~Senay, D.~Allured, S.~Leroux, and C.~Funk, 2012: Seasonality of african precipitation from 1996 to 2009. \textit{Journal of Climate}, \textbf{25~(12)}, 4304--4322, \doi{10.1175/jcli-d-11-00157.1}.

\bibitem[{Liebmann et~al.(2014)}]{Liebmann_2014}
Liebmann, B., and Coauthors, 2014: Understanding recent eastern horn of africa rainfall variability and change. \textit{Journal of Climate}, \textbf{27~(23)}, 8630--8645, \doi{10.1175/jcli-d-13-00714.1}.

\bibitem[{Lisonbee et~al.(2021)Lisonbee, Woloszyn,, and Skumanich}]{Lisonbee_2021}
Lisonbee, J., M.~Woloszyn, and M.~Skumanich, 2021: Making sense of flash drought: definitions, indicators, and where we go from here. \textit{Journal of Applied and Service Climatology}, \textbf{2021~(1)}, 1--19, \doi{10.46275/joasc.2021.02.001}.

\bibitem[{Lyapustin(2023)}]{Lyapustin_2023_MODIS}
Lyapustin, A., 2023: Modis/terra+aqua vegetation index from maiac, daily l3 global 0.05deg cmg v061. NASA EOSDIS Land Processes Distributed Active Archive Center, \urlprefix\url{https://lpdaac.usgs.gov/products/mcd19a3cmgv061/}, \doi{10.5067/MODIS/MCD19A3CMG.061}.

\bibitem[{Lyon(2014)}]{Lyon_2014}
Lyon, B., 2014: Seasonal drought in the greater horn of africa and its recent increase during the march--may long rains. \textit{Journal of Climate}, \textbf{27~(21)}, 7953--7975, \doi{10.1175/jcli-d-13-00459.1}.

\bibitem[{Macharia et~al.(2022)Macharia, Fankhauser, Selker, Neff,, and Thomas}]{Macharia_2022}
Macharia, D., K.~Fankhauser, J.~S. Selker, J.~C. Neff, and E.~A. Thomas, 2022: Validation and intercomparison of satellite-based rainfall products over africa with tahmo in-situ rainfall observations. \textit{Journal of Hydrometeorology}, \doi{10.1175/jhm-d-21-0161.1}.

\bibitem[{Madakumbura et~al.(2021)Madakumbura, Thackeray, Norris, Goldenson,, and Hall}]{Madakumbura_2021}
Madakumbura, G.~D., C.~W. Thackeray, J.~Norris, N.~Goldenson, and A.~Hall, 2021: Anthropogenic influence on extreme precipitation over global land areas seen in multiple observational datasets. \textit{Nature Communications}, \textbf{12~(1)}, \doi{10.1038/s41467-021-24262-x}.

\bibitem[{Maidment et~al.(2014)Maidment, Grimes, Allan, Tarnavsky, Stringer, Hewison, Roebeling,, and Black}]{Maidment_2014}
Maidment, R.~I., D.~Grimes, R.~P. Allan, E.~Tarnavsky, M.~Stringer, T.~Hewison, R.~Roebeling, and E.~Black, 2014: The 30 year tamsat african rainfall climatology and time series (tarcat) data set. \textit{Journal of Geophysical Research: Atmospheres}, \textbf{119~(18)}, \doi{10.1002/2014jd021927}.

\bibitem[{Maidment et~al.(2017)}]{Maidment_2017}
Maidment, R.~I., and Coauthors, 2017: A new, long-term daily satellite-based rainfall dataset for operational monitoring in africa. \textit{Scientific Data}, \textbf{4~(1)}, \doi{10.1038/sdata.2017.63}.

\bibitem[{Marcolongo et~al.(2022)Marcolongo, Vladymyrov, Lienert, Peleg, Haug,, and Zscheischler}]{Marcolongo_2022}
Marcolongo, A., M.~Vladymyrov, S.~Lienert, N.~Peleg, S.~Haug, and J.~Zscheischler, 2022: Predicting years with extremely low gross primary production from daily weather data using convolutional neural networks. \textit{Environmental Data Science}, \textbf{1}, \doi{10.1017/eds.2022.1}.

\bibitem[{Misra(2003)}]{Misra_2003}
Misra, V., 2003: The influence of pacific sst variability on the precipitation over southern africa. \textit{Journal of Climate}, \textbf{16~(14)}, 2408--2418, \doi{10.1175/2785.1}.

\bibitem[{Mukherjee and Mishra(2022{\natexlab{a}})Mukherjee, and Mishra}]{Mukherjee_2022a}
Mukherjee, S., and A.~K. Mishra, 2022{\natexlab{a}}: Global flash drought analysis: Uncertainties from indicators and datasets. \textit{Earth's Future}, \textbf{10~(6)}, \doi{10.1029/2022ef002660}.

\bibitem[{Mukherjee and Mishra(2022{\natexlab{b}})Mukherjee, and Mishra}]{Mukherjee_2022b}
Mukherjee, S., and A.~K. Mishra, 2022{\natexlab{b}}: A multivariate flash drought indicator for identifying global hotspots and associated climate controls. \textit{Geophysical Research Letters}, \textbf{49~(2)}, \doi{10.1029/2021gl096804}.

\bibitem[{Myneni et~al.(2021)Myneni, Knyazikhin,, and Park}]{Myneni_2021_MODIS}
Myneni, R., Y.~Knyazikhin, and T.~Park, 2021: Modis/terra+aqua leaf area index/fpar 8-day l4 global 500m sin grid v061. NASA EOSDIS Land Processes Distributed Active Archive Center, \urlprefix\url{https://lpdaac.usgs.gov/products/mcd15a2hv061/}, \doi{10.5067/MODIS/MCD15A2H.061}.

\bibitem[{Nafii et~al.(2022)Nafii, Taleb, El~Mesbahi, Ezzaouini,, and El~Bilali}]{Nafii_2022}
Nafii, A., A.~Taleb, M.~El~Mesbahi, M.~A. Ezzaouini, and A.~El~Bilali, 2022: Early forecasting hydrological and agricultural droughts in the bouregreg basin using a machine learning approach. \textit{Water}, \textbf{15~(1)}, 122, \doi{10.3390/w15010122}.

\bibitem[{Naumann et~al.(2012)Naumann, Barbosa, Carrao, Singleton,, and Vogt}]{Naumann_2012}
Naumann, G., P.~Barbosa, H.~Carrao, A.~Singleton, and J.~Vogt, 2012: Monitoring drought conditions and their uncertainties in africa using trmm data. \textit{Journal of Applied Meteorology and Climatology}, \textbf{51~(10)}, 1867--1874, \doi{10.1175/jamc-d-12-0113.1}.

\bibitem[{Nicholson(2015)}]{Nicholson_2015}
Nicholson, S.~E., 2015: Long‐term variability of the east african `short rains' and its links to large‐scale factors. \textit{International Journal of Climatology}, \textbf{35~(13)}, 3979--3990, \doi{10.1002/joc.4259}.

\bibitem[{Nicholson(2017)}]{Nicholson_2017}
Nicholson, S.~E., 2017: Climate and climatic variability of rainfall over eastern africa. \textit{Reviews of Geophysics}, \textbf{55~(3)}, 590--635, \doi{10.1002/2016rg000544}.

\bibitem[{NicholsonI and Kim(1997)NicholsonI, and Kim}]{Nicholson_1997}
NicholsonI, S.~E., and J.~Kim, 1997: The relationship of the el ni{\~n}o--southern oscillation to african rainfall. \textit{International Journal of Climatology}, \textbf{17~(2)}, 117--135, \doi{10.1002/(sici)1097-0088(199702)17:2<117::aid-joc84>3.0.co;2-o}.

\bibitem[{Nipen et~al.(2024)}]{Nipen_2024}
Nipen, T.~N., and Coauthors, 2024: Regional data-driven weather modeling with a global stretched-grid. arXiv, \urlprefix\url{https://arxiv.org/abs/2409.02891}, \doi{10.48550/ARXIV.2409.02891}.

\bibitem[{Noguera et~al.(2021)Noguera, Dom{\'\i}nguez-Castro,, and Vicente-Serrano}]{Noguera_2021}
Noguera, I., F.~Dom{\'\i}nguez-Castro, and S.~M. Vicente-Serrano, 2021: Flash drought response to precipitation and atmospheric evaporative demand in spain. \textit{Atmosphere}, \textbf{12~(2)}, 165, \doi{10.3390/atmos12020165}.

\bibitem[{Olaniyan et~al.(2025)Olaniyan, Woolnough, Andrade, Hirons, Thompson,, and Lawal}]{Olaniyan_2025}
Olaniyan, E.~A., S.~J. Woolnough, F.~M.~D. Andrade, L.~C. Hirons, E.~Thompson, and K.~A. Lawal, 2025: Performance evaluation of real-time sub-to-seasonal (s2s) rainfall forecasts over west africa of 2020 and 2021 monsoon seasons for operational use. \textit{Atmosphere}, \textbf{16~(9)}, 1072, \doi{10.3390/atmos16091072}.

\bibitem[{Olson et~al.(2019)Olson, Bolvin,, and Huffman}]{Olson_2019-IMERG_land-sea_data}
Olson, W., D.~Bolvin, and G.~Huffman, 2019: Land/sea static mask relevant to imerg precipitation 0.1x0.1 degree v2. NASA Goddard Earth Sciences Data and Information Services Center, \urlprefix\url{https://disc.gsfc.nasa.gov/datacollection/GPM_IMERG_LandSeaMask_2.html}, \doi{10.5067/6P5EM1HPR3VD}.

\bibitem[{Osman et~al.(2022)}]{Osman_2022}
Osman, M., and Coauthors, 2022: Diagnostic classification of flash drought events reveals distinct classes of forcings and impacts. \textit{Journal of Hydrometeorology}, \textbf{23~(2)}, 275--289, \doi{10.1175/jhm-d-21-0134.1}.

\bibitem[{Otkin et~al.(2018)Otkin, Svoboda, Hunt, Ford, Anderson, Hain,, and Basara}]{Otkin_2018}
Otkin, J.~A., M.~Svoboda, E.~D. Hunt, T.~W. Ford, M.~C. Anderson, C.~Hain, and J.~B. Basara, 2018: Flash droughts: A review and assessment of the challenges imposed by rapid-onset droughts in the united states. \textit{Bulletin of the American Meteorological Society}, \textbf{99~(5)}, 911--919, \doi{10.1175/bams-d-17-0149.1}.

\bibitem[{Otkin et~al.(2021)}]{Otkin_2021}
Otkin, J.~A., and Coauthors, 2021: Development of a flash drought intensity index. \textit{Atmosphere}, \textbf{12~(6)}, 741, \doi{10.3390/atmos12060741}.

\bibitem[{Palmer et~al.(2023)}]{Palmer_2023}
Palmer, P.~I., and Coauthors, 2023: Drivers and impacts of eastern african rainfall variability. \textit{Nature Reviews Earth \& Environment}, \textbf{4~(4)}, 254--270, \doi{10.1038/s43017-023-00397-x}.

\bibitem[{Phakula et~al.(2024)Phakula, Landman,, and Engelbrecht}]{Phakula_2024}
Phakula, S., W.~A. Landman, and C.~J. Engelbrecht, 2024: Literature survey of <scp>subseasonal‐to‐seasonal</scp> predictions in the southern hemisphere. \textit{Meteorological Applications}, \textbf{31~(1)}, \doi{10.1002/met.2170}.

\bibitem[{{Precipitation Processing System (PPS) At NASA GSFC}(2023)}]{Huffman_2023-IMERG_data}
{Precipitation Processing System (PPS) At NASA GSFC}, 2023: Gpm imerg final precipitation l3 1 day 0.1 degree x 0.1 degree v07. NASA Goddard Earth Sciences Data and Information Services Center, \urlprefix\url{https://disc.gsfc.nasa.gov/datacollection/GPM_3IMERGDF_07.html}, \doi{10.5067/GPM/IMERGDF/DAY/07}.

\bibitem[{Rakkasagi et~al.(2023)Rakkasagi, Poonia,, and Goyal}]{Rakkasagi_2023}
Rakkasagi, S., V.~Poonia, and M.~K. Goyal, 2023: Flash drought as a new climate threat: drought indices, insights from a study in india and implications for future research. \textit{Journal of Water and Climate Change}, \textbf{14~(9)}, 3368--3384, \doi{10.2166/wcc.2023.347}.

\bibitem[{Rayner et~al.(2003)Rayner, Parker, Horton, Folland, Alexander, Rowell, Kent,, and Kaplan}]{Rayner_2003_hadisst}
Rayner, N.~A., D.~E. Parker, E.~B. Horton, C.~K. Folland, L.~V. Alexander, D.~P. Rowell, E.~C. Kent, and A.~Kaplan, 2003: Global analyses of sea surface temperature, sea ice, and night marine air temperature since the late nineteenth century. \textit{Journal of Geophysical Research: Atmospheres}, \textbf{108~(D14)}, \doi{10.1029/2002jd002670}.

\bibitem[{Rodell et~al.(2004)}]{Rodell_2004_GLDAS}
Rodell, M., and Coauthors, 2004: The global land data assimilation system. \textit{Bulletin of the American Meteorological Society}, \textbf{85~(3)}, 381--394, \doi{10.1175/bams-85-3-381}.

\bibitem[{Running et~al.(2021)Running, Mu, Zhao,, and Moreno}]{Running_2021_MODIS}
Running, S., Q.~Mu, M.~Zhao, and A.~Moreno, 2021: Modis/terra net evapotranspiration gap-filled 8-day l4 global 500m sin grid v061. NASA EOSDIS Land Processes Distributed Active Archive Center, \urlprefix\url{https://lpdaac.usgs.gov/products/mod16a2gfv061/}, \doi{10.5067/MODIS/MOD16A2GF.061}.

\bibitem[{Saji and Yamagata(2003)Saji, and Yamagata}]{Saji_2003}
Saji, N.~H., and T.~Yamagata, 2003: Possible impacts of indian ocean dipole mode events on global climate. \textit{Climate Research}, \textbf{25~(2)}, 151--169.

\bibitem[{Scaife and Smith(2018)Scaife, and Smith}]{Scaife_2018}
Scaife, A.~A., and D.~Smith, 2018: A signal-to-noise paradox in climate science. \textit{npj Climate and Atmospheric Science}, \textbf{1~(1)}, \doi{10.1038/s41612-018-0038-4}.

\bibitem[{Schreck et~al.(2025{\natexlab{a}})}]{Schreck_2025_CREDIT}
Schreck, J.~S., and Coauthors, 2025{\natexlab{a}}: Community research earth digital intelligence twin: a scalable framework for ai-driven earth system modeling. \textit{npj Climate and Atmospheric Science}, \textbf{8~(1)}, \doi{10.1038/s41612-025-01125-6}.

\bibitem[{Schreck et~al.(2025{\natexlab{b}})}]{Schreck_2025}
Schreck, J.~S., and Coauthors, 2025{\natexlab{b}}: Controllable probabilistic forecasting with stochastic decomposition layers. arXiv, \urlprefix\url{https://arxiv.org/abs/2512.18815}, \doi{10.48550/ARXIV.2512.18815}.

\bibitem[{Sehgal et~al.(2021)Sehgal, Gaur,, and Mohanty}]{Sehgal_2021}
Sehgal, V., N.~Gaur, and B.~P. Mohanty, 2021: Global flash drought monitoring using surface soil moisture. \textit{Water Resources Research}, \textbf{57~(9)}, \doi{10.1029/2021wr029901}.

\bibitem[{Sha et~al.(2025)Sha, Schreck, Chapman,, and Gagne}]{Sha_2025}
Sha, Y., J.~S. Schreck, W.~Chapman, and D.~J. Gagne, 2025: Improving ai weather prediction models using global mass and energy conservation schemes. \textit{Journal of Advances in Modeling Earth Systems}, \textbf{17~(11)}, \doi{10.1029/2025ms005138}.

\bibitem[{Sheffield et~al.(2014)}]{Sheffield_2014}
Sheffield, J., and Coauthors, 2014: A drought monitoring and forecasting system for sub-sahara african water resources and food security. \textit{Bulletin of the American Meteorological Society}, \textbf{95~(6)}, 861--882, \doi{10.1175/bams-d-12-00124.1}.

\bibitem[{Shi et~al.(2016)Shi, Caballero, Husz{\'a}r, Totz, Aitken, Bishop, Rueckert,, and Wang}]{Shi_2016}
Shi, W., J.~Caballero, F.~Husz{\'a}r, J.~Totz, A.~P. Aitken, R.~Bishop, D.~Rueckert, and Z.~Wang, 2016: Real-time single image and video super-resolution using an efficient sub-pixel convolutional neural network. arXiv, \urlprefix\url{https://arxiv.org/abs/1609.05158}, \doi{10.48550/ARXIV.1609.05158}.

\bibitem[{Shukla et~al.(2014)Shukla, McNally, Husak,, and Funk}]{Shukla_2014}
Shukla, S., A.~McNally, G.~Husak, and C.~Funk, 2014: A seasonal agricultural drought forecast system for food-insecure regions of east africa. \textit{Hydrology and Earth System Sciences}, \textbf{18~(10)}, 3907--3921, \doi{10.5194/hess-18-3907-2014}.

\bibitem[{Tarnavsky et~al.(2014)Tarnavsky, Grimes, Maidment, Black, Allan, Stringer, Chadwick,, and Kayitakire}]{Tarnavsky_2014}
Tarnavsky, E., D.~Grimes, R.~Maidment, E.~Black, R.~P. Allan, M.~Stringer, R.~Chadwick, and F.~Kayitakire, 2014: Extension of the tamsat satellite-based rainfall monitoring over africa and from 1983 to present. \textit{Journal of Applied Meteorology and Climatology}, \textbf{53~(12)}, 2805--2822, \doi{10.1175/jamc-d-14-0016.1}.

\bibitem[{Thomas et~al.(2020)}]{Thomas_2020b_DRIP}
Thomas, E.~A., and Coauthors, 2020: The drought resilience impact platform (drip): Improving water security through actionable water management insights. \textit{Frontiers in Climate}, \textbf{2}, \doi{10.3389/fclim.2020.00006}.

\bibitem[{Tierney et~al.(2013)Tierney, Smerdon, Anchukaitis,, and Seager}]{Tierney_2013}
Tierney, J.~E., J.~E. Smerdon, K.~J. Anchukaitis, and R.~Seager, 2013: Multidecadal variability in east african hydroclimate controlled by the indian ocean. \textit{Nature}, \textbf{493~(7432)}, 389--392, \doi{10.1038/nature11785}.

\bibitem[{Tyagi et~al.(2022)Tyagi, Zhang, Saraswat, Sahany, Mishra,, and Niyogi}]{Tyagi_2022}
Tyagi, S., X.~Zhang, D.~Saraswat, S.~Sahany, S.~K. Mishra, and D.~Niyogi, 2022: Flash drought: Review of concept, prediction and the potential for machine learning, deep learning methods. \textit{Earth{\textquotesingle}s Future}, \doi{10.1029/2022ef002723}.

\bibitem[{UNDRR(2024)}]{undrr_2024}
UNDRR, 2024: Horn of africa floods and drought, 2020-2023 - forensic analysis. \urlprefix\url{https://www.undrr.org/resource/horn-africa-floods-and-drought-2020-2023-forensic-analysis}.

\bibitem[{Verdin et~al.(2005)Verdin, Funk, Senay,, and Choularton}]{Verdin_2005}
Verdin, J., C.~Funk, G.~Senay, and R.~Choularton, 2005: Climate science and famine early warning. \textit{Philosophical Transactions of the Royal Society B: Biological Sciences}, \textbf{360~(1463)}, 2155--2168, \doi{10.1098/rstb.2005.1754}.

\bibitem[{Vicente-Serrano et~al.(2012)}]{Vicente_Serrano_2012}
Vicente-Serrano, S.~M., and Coauthors, 2012: Challenges for drought mitigation in africa: The potential use of geospatial data and drought information systems. \textit{Applied Geography}, \textbf{34}, 471--486, \doi{10.1016/j.apgeog.2012.02.001}.

\bibitem[{Wainwright et~al.(2020)Wainwright, Finney, Kilavi, Black,, and Marsham}]{Wainwright_2020}
Wainwright, C.~M., D.~L. Finney, M.~Kilavi, E.~Black, and J.~H. Marsham, 2020: Extreme rainfall in east africa, october 2019--january 2020 and context under future climate change. \textit{Weather}, \textbf{76~(1)}, 26--31, \doi{10.1002/wea.3824}.

\bibitem[{Wainwright et~al.(2021)Wainwright, Marsham, Rowell, Finney,, and Black}]{Wainwright_2021}
Wainwright, C.~M., J.~H. Marsham, D.~P. Rowell, D.~L. Finney, and E.~Black, 2021: Future changes in seasonality in east africa from regional simulations with explicit and parameterized convection. \textit{Journal of Climate}, \textbf{34~(4)}, 1367--1385, \doi{10.1175/jcli-d-20-0450.1}.

\bibitem[{Wang et~al.(2018)Wang, Chen, Du, Zhang, Xia,, and Deng}]{Wang_2018}
Wang, J., J.~Chen, J.~Du, Y.~Zhang, Y.~Xia, and G.~Deng, 2018: Sensitivity of ensemble forecast verification to model bias. \textit{Monthly Weather Review}, \textbf{146~(3)}, 781--796, \doi{10.1175/mwr-d-17-0223.1}.

\bibitem[{Watt-Meyer et~al.(2024)}]{Watt-Meyer_2024}
Watt-Meyer, O., and Coauthors, 2024: Ace2: Accurately learning subseasonal to decadal atmospheric variability and forced responses. arXiv, \urlprefix\url{https://arxiv.org/abs/2411.11268}, \doi{10.48550/ARXIV.2411.11268}.

\bibitem[{Wu et~al.(2021)Wu, Su, Singh, Feng,, and Niu}]{Wu_2021}
Wu, H., X.~Su, V.~P. Singh, K.~Feng, and J.~Niu, 2021: Agricultural drought prediction based on conditional distributions of vine copulas. \textit{Water Resources Research}, \textbf{57~(8)}, \doi{10.1029/2021wr029562}.

\bibitem[{Yuan et~al.(2019)Yuan, Wang, Wu, Ji, Sheffield,, and Zhang}]{Yuan_2019}
Yuan, X., L.~Wang, P.~Wu, P.~Ji, J.~Sheffield, and M.~Zhang, 2019: Anthropogenic shift towards higher risk of flash drought over china. \textit{Nature Communications}, \textbf{10~(1)}, \doi{10.1038/s41467-019-12692-7}.

\bibitem[{Yuan et~al.(2023)Yuan, Wang, Ji, Wu, Sheffield,, and Otkin}]{Yuan_2023}
Yuan, X., Y.~Wang, P.~Ji, P.~Wu, J.~Sheffield, and J.~A. Otkin, 2023: A global transition to flash droughts under climate change. \textit{Science}, \textbf{380~(6641)}, 187--191, \doi{10.1126/science.abn6301}.

\bibitem[{Zhang et~al.(2022)}]{Zhang_2022}
Zhang, L., and Coauthors, 2022: Analysis of flash droughts in china using machine learning. \textit{Hydrology and Earth System Sciences}, \textbf{26~(12)}, 3241--3261, \doi{10.5194/hess-26-3241-2022}.

\end{thebibliography}

\end{document}